\documentclass[journal]{IEEEtran}
\ifCLASSINFOpdf
\else
\fi
\usepackage{booktabs}
\usepackage{graphicx} 
\usepackage{stfloats}

\usepackage{textcomp}
\usepackage{stfloats}
\usepackage{verbatim}
\usepackage{graphicx}
\usepackage{booktabs}
\usepackage{amsmath,amsfonts}
\usepackage{algorithmic}
\usepackage{array}
\usepackage{stfloats}
\usepackage{verbatim}
\usepackage{graphicx}
\usepackage{balance}
\usepackage{balance}
\usepackage[utf8]{inputenc}
\usepackage{caption}
\usepackage{graphicx}
\usepackage{float} 
\usepackage{subcaption}
\usepackage{color}
\usepackage{makecell}

\newcommand{\specialcell}[2][c]{%
	\begin{tabular}[#1]{@{}l@{}}#2\end{tabular}}

\begin{document}
\title{WiFi-Diffusion: Achieving Fine-Grained WiFi Radio Map Estimation with Ultra-Low Sampling Rate by Diffusion Models}
\author{Zhiyuan~Liu,~\IEEEmembership{Student Member,~IEEE, }
Shuhang~Zhang,~\IEEEmembership{Member,~IEEE, }
Qingyu~Liu,~\IEEEmembership{Member,~IEEE, }
Hongliang~Zhang,~\IEEEmembership{Member,~IEEE, } Lingyang~Song,~\IEEEmembership{Fellow,~IEEE}
\thanks{Received 30 November 2024; revised 15 April 2025.
This work was supported in part by the Natural Science Foundation of China under Grant 62401025; in part by the Foundation for Basic and Applied Research of Guangdong Province under the Youth Fund Project of the Guangdong-Shenzhen Joint Fund under Grant 2023A1515110120.}
\thanks{Zhiyuan Liu is with the School of Electronic and Computer Engineering,  Peking University Shenzhen Graduate School, Shenzhen, China, 518055 (e-mail: 
{liuzhiyuan@stu.pku.edu.cn}).}
\thanks{Shuhang Zhang is with the Pengcheng Laboratory, Shenzhen, China, 518055 (email: {zhangshh01@pcl.ac.cn}).}
\thanks{Qingyu Liu is with the School of Electronic and Computer Engineering,  Peking University Shenzhen Graduate School, Shenzhen, China, 518055, and also with the Pengcheng Laboratory, Shenzhen, China, 518055 (e-mail: {qy.liu@pku.edu.cn}).}
\thanks{Hongliang Zhang is with the Department of Electronics, Peking University, Beijing, China, 100871 (email: 
{hongliang.zhang@pku.edu.cn}).}
\thanks{Lingyang Song is with the Department of Electronics, Peking University, Beijing, China, 100871, with the School of Electronic and Computer Engineering, Peking University Shenzhen Graduate School, Shenzhen,
China, 518055, and also with the Pengcheng Laboratory, Shenzhen, China, 518055 (e-mail: 
{lingyang.song@pku.edu.cn}).}}


\maketitle
\thispagestyle{empty}
\pagestyle{empty}

\begin{abstract}
The radio map presents communication parameters of interest, e.g., received signal strength, at every point across a geographical region. 
It can be leveraged to improve the efficiency of spectrum utilization in the region, particularly critical for unlicensed WiFi spectrum.
The problem of fine-grained radio map estimation is to utilize radio samples collected by sensors sparsely distributed in the region to infer a high-resolution radio map.
This problem is challenging due to the ultra-low sampling rate, i.e., because the number of available samples is far fewer than the high resolution required for radio map estimation.
We propose WiFi-Diffusion -- a novel generative framework for achieving fine-grained WiFi radio map estimation using diffusion models.
WiFi-Diffusion employs the creative power of generative AI to address the ultra-low sampling rate challenge and consists of three blocks:
1) a boost block, using prior information such as the layout of obstacles to optimize the diffusion model;
2) a generation block, leveraging the diffusion model to generate a candidate set of fine-grained radio maps;
and 3) an election block, utilizing the radio propagation model as a guide to find the best fine-grained radio map from the candidate set.
Extensive simulations demonstrate that 1) the fine-grained radio map generated by WiFi-Diffusion is ten times better than those produced by state-of-the-art (SOTA) when they use the same ultra-low sampling rate; and 2) WiFi-Diffusion achieves comparable fine-grained radio map quality with only one-fifth of the sampling rate required by SOTA.
\end{abstract}


\section{Introduction}

The radio map depicts parameters of interest in communication channels, such as the received signal strength (RSS), at every point of a certain geographical region~\cite{RME}.
It can significantly improve the performance of many wireless applications, enabling functions such as dynamic spectrum access \cite{dynamic-spectrum-access}, spectrum sharing~\cite{RadioGAT}, and interference management~\cite{smartWiFi}.
The fine-grained radio map is a high-resolution radio map for a large geographical region.
It is particularly critical for WiFi, which operates in unlicensed spectra and plays a pivotal role in shaping the future of wireless connectivity.
Currently, the WiFi spectra are extensively utilized by numerous devices, transmitting a vast amount of data.
It has been estimated that almost $70\%$ of global Internet traffic crosses a WiFi network \cite{forecast2019cisco}.
With the anticipated rapid growth in WiFi traffic, advanced techniques in radio recognition, spectrum management, and network resource allocation are needed to ensure fast, reliable, and ubiquitous connectivity enabled by WiFi.
The fine-grained radio map can serve as an indispensable foundation for the efficient utilization of WiFi in a large area.

Radio map estimation problems can generally be classified into two categories: sampling-free problem and sampling-based problem. 
The sampling-free problem uses the information from the transmitters to estimate radio maps. 
Existing studies solving these problems include~\cite{ray-tracing,dominant-path,eg2,eg}.
The sampling-free problem assumes knowledge of the information (e.g., number, locations, and transmitting power) of transmitters.
This assumption is strong and may not hold in practice, especially for WiFi that operates in unlicensed spectra, where there often exist many transmitters unknown to users.
In contrast, the sampling-based problem uses certain prior information, e.g., the layout of obstacles and the limited number of radio samples collected by sensors sparsely distributed in the geographical region, to predict radio maps.
In this paper, we focus on the sampling-based fine-grained radio map estimation problem for the WiFi spectrum.

\begin{figure*}[th]
	\centering
	\includegraphics[width=0.8\linewidth]{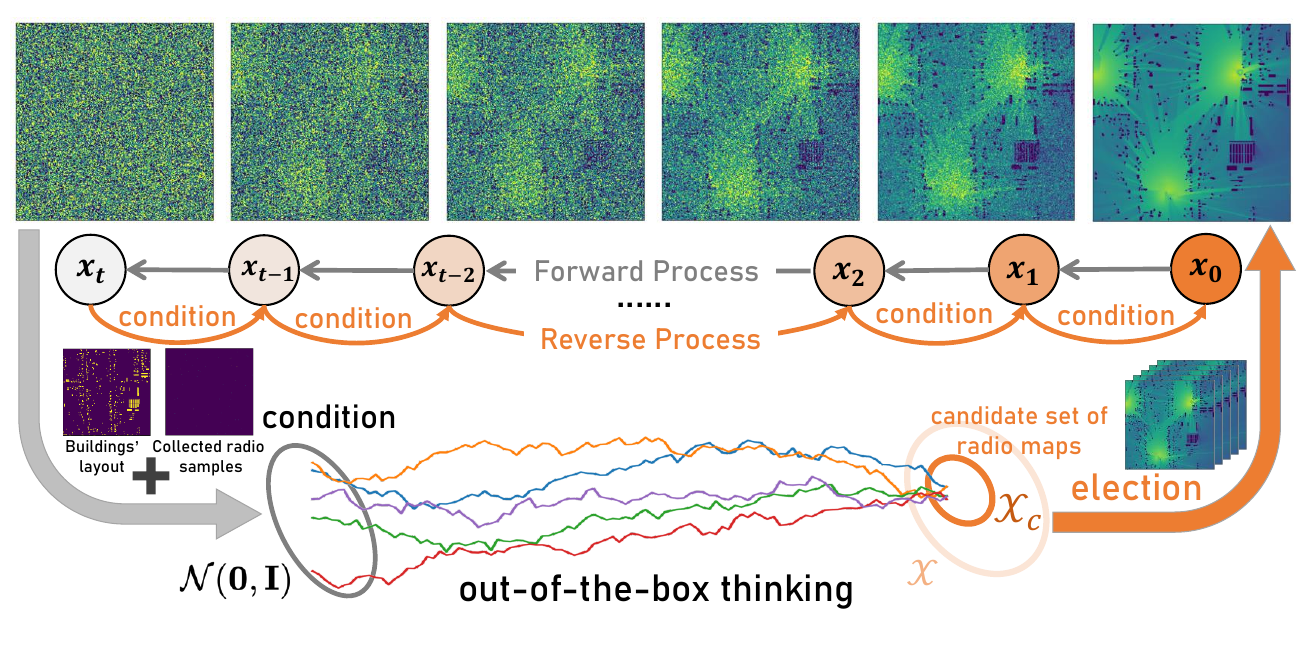}
    \caption{WiFi-Diffusion--a diffusion model-based framework for fine-grained radio map estimation with ultra-low sampling rate.}
	\label{diffMethod}
\end{figure*}

Approaches for solving the sampling-based radio map estimation problem can generally be classified into two categories: interpolation-based approaches and deep learning-based approaches.
Interpolation-based approaches utilize mathematical and statistical methodologies to estimate radio maps.
Well-known methods include radial basis functions (RBF)~\cite{rbf}, spline~\cite{spline}, and ordinary kriging~\cite{kriging}.
However, these methods often fail to construct high-quality radio maps for real-world environments that are dynamic and geographically complex.
This is because they rely on mathematical calculations and do not account for certain critical prior information, such as the layout of obstacles, which can significantly impact radio signal propagation.

Recently, cutting-edge deep learning techniques have been applied to the problem of radio map estimation, surpassing the performance of traditional interpolation-based approaches.
{Levie \emph{et al.} \cite{RadioUNet} proposed a U-Net-based scheme for estimating radio maps.}
{Teganya \emph{et al.} \cite{AE} and Locke \emph{et al.} \cite{autoencoder} designed autoencoder-based methods for constructing radio maps.}
{He \emph{et al.} \cite{ResNet} designed a ResNet-based approach for radio map estimation.}
Other deep learning approaches, e.g., Feedforward Neural Networks~\cite{FEED_ref}, Generative Adversarial networks~\cite{GAN_ref}, and Graph Attention Networks~\cite{RadioGAT}, were used for radio map estimation, too.

The fine-grained radio map estimation problem is challenging due to the \emph{ultra-low sampling rate}, i.e., because of the inherent conflict between the high resolution required for radio map estimation and the ultra-small number of radio samples collected by sensors.
In practice, the price of sensors that can collect radio samples is high.
For example, even a low-end sensor like the Ettus USRP B200mini can cost more than $1,000$ USD.
To generate a radio map with a resolution of $256 \times 256$ like those in~\cite{RadioUNet} and~\cite{AE}, even a sampling rate of $1\%$ would require over $600$ sensors, resulting in a total cost exceeding $600,000$ USD, which is impractical.
Therefore, it is crucial for radio map estimation approaches to be capable of producing fine-grained, high-quality radio maps at ultra-low sampling rates (e.g., less than $0.1\%$).
However, existing deep learning-based approaches, such as those in~\cite{autoencoder,AE,ResNet,FEED_ref,GAN_ref,RadioUNet,RadioGAT}, have sampling rates ranging from $1\%$ to $10\%$.
They all fail to construct fine-grained, high-quality radio maps at ultra-low sampling rates.

In order to deal with the ultra-low sampling rate challenge, in this paper, we leverage advanced diffusion-based generative techniques for radio map estimation.
One key advantage of generative models compared to traditional discriminative models is their ``out-of-the-box thinking'' ability to generate a wide range of diverse predictions from limited inputs. 
This creative power makes generative models well-suited for estimating radio maps at ultra-low sampling rates.
Among the most prominent and powerful generative models is the diffusion model, which was first introduced in~\cite{beforeDDPM} and significantly improved in~\cite{DDPM}. Since then, it has been successfully applied to various tasks such as image synthesis~\cite{diffusionIMG}, natural language processing~\cite{nlp}, and molecule design~\cite{Protein}.
To the best of our knowledge, we are the first to use diffusion-based generative techniques to solve the problem of radio map estimation.
Specifically, we design WiFi-Diffusion, a diffusion model-based algorithm capable of generating fine-grained, high-quality radio maps at ultra-low sampling rates.

As shown in Fig.~\ref{diffMethod}, the key idea of WiFi-Diffusion is first to use the diffusion model to generate a candidate set of fine-grained radio maps and then to find the best map from the set as the final output.
To optimize the quality of the final output, WiFi-Diffusion solves the following two critical problems:
\begin{itemize}
    \item How to enhance the qualities of radio maps in the candidate set?
    \item How to find the best map from the candidate set?
\end{itemize}

To solve the first problem, WiFi-Diffusion incorporates a \emph{boost block}, where it leverages prior information, including the layout of obstacles, the law of radio propagation, and collected radio samples, as inputs to a UNet model with self-attention.
The UNet model identifies ``critical points" within the geographical region from the perspective of constructing radio maps and conveys the information of these points to the diffusion model.
Subsequently, WiFi-Diffusion employs the diffusion model to generate a set of fine-grained radio map candidates from the limited samples during a \emph{generation block}.
Although the qualities of candidate maps can be enhanced by the previous boost block, due to the ultra-low sampling rate, the majority of these maps are poor, with only a minority being accurate. 
Then WiFi-Diffusion proceeds to an \emph{election block} that solves the second problem and selects the best map from the candidate set.
In this block, WiFi-Diffusion uses the mathematical radio propagation model as a guide to identify the map that most closely adheres to the propagation model in the candidate set and selects it as the final output.
Through this novel generative framework of three blocks, WiFi-Diffusion is able to generate fine-grained, high-quality radio maps at ultra-low sampling rates.

Specifically, the main contributions of this paper are summarized as follows:

\begin{itemize}
\item We propose WiFi-Diffusion -- a novel generative framework for fine-grained radio map estimation for the WiFi spectrum.
To address the ultra-low sampling rate challenge of fine-grained radio map estimation, WiFi-Diffusion leverages the creative power of diffusion-based generative models to produce a diverse set of fine-grained radio map candidates.
Subsequently, WiFi-Diffusion identifies and selects the best map from this set as the final output.
To the best of our knowledge, we are the first to use diffusion-based generative techniques to solve the problem of sampling-based radio map estimation.

\item Although a diverse set of fine-grained radio maps can be generated by the diffusion model, the majority of them are quite poor and only a minority are accurate.
To guarantee a high-quality final output, WiFi-Diffusion incorporates a boost block prior to generating fine-grained radio map candidates and an election block following their generation.
In the boost block, WiFi-Diffusion utilizes prior information, e.g., layout of obstacles, principle of radio propagation, and collected radio samples, to enhance the qualities of fine-grained radio map candidates generated by the diffusion model.
In the election block, WiFi-Diffusion employs the mathematical radio propagation model as a guide to find the best map from the diverse candidates generated by the diffusion model.
With the boost block, generation block, and election block, WiFi-Diffusion is able to generate fine-grained, high-quality radio maps at ultra-low sampling rates.

\item Extensive simulations based on the 5750 MHz (WiFi 5G) BART-Lab radio map dataset show that WiFi-Diffusion is effective.
Specifically, simulations demonstrate that 
1) WiFi-Diffusion generates fine-grained, high-quality radio maps when the sampling rate is less than $0.1\%$;
2) the fine-grained radio map generated by WiFi-Diffusion is ten times better than that generated by state-of-the-art (SOTA) when they share the same ultra-low sampling rate;
3) WiFi-Diffusion requires only one-fifth of the sampling rate needed by SOTA to generate comparable fine-grained radio maps.
\end{itemize}

The rest of the paper is organized as follows. 
In Section~\ref{system-model-and-problem-description}, we describe the system model and introduce the radio map estimation problem. 
In Section~\ref{scheme},
we present the key ideas of the design of WiFi-Diffusion. 
In Section~\ref{scheme-detail}, we offer the design details of WiFi-Diffusion. 
In Section~\ref{Experimental}, we perform extensive simulations to evaluate the performance of WiFi-Diffusion.
{Section~\ref{sec:app} discusses potential applications of WiFi-Diffusion.}
Section~\ref{Conclusion} concludes the paper.

\section{Problem Description} \label{system-model-and-problem-description}

\begin{table}[]
	\centering
	\caption{Summary of key notations.}
	\renewcommand{\arraystretch}{1.3}
	\begin{tabular}{cp{2.2in}}
		\hline
		Symbol & Definition \\ \hline
        $\mathcal{Y}$& region of the radio map\\
        $k$& number of radio samples collected by sensors\\
        $\mathcal{S}$& set of $k$ sensors that can collect radio samples\\
        $H$& number of grids (points) in one row in one radio map\\
        $W$& number of grids (points) in one column in one radio map\\	 
         $\mathbf{S}_k$& an $H\times W$ matrix, where the element at index $(h_s,w_s)$ is 0 if there is no sensor located at that grid point; otherwise, it is the RSS value of the sample collected by the sensor\\
   $\mathbf{B}$& layout of buildings \\
   
   \hline
\end{tabular}
\label{Notations}
\end{table}

Consider a two-dimensional region of interest $\mathcal{Y}\subset\mathbb{R}^2$. 
Consider a set of sensors $\mathcal{S}=\{1,2,\cdots,k\}$ (the number of sensors is $k$) where each sensor $s\in\mathcal{S}$ is located at a unique coordinate $y_s\in\mathcal{Y}$.
Each sensor $s$ measures the RSS in a given WiFi frequency band $\mathcal{F}$ at its location $y_s$, where $r_s(f)$ denotes the RSS in frequency $f\in\mathcal{F}$ (assuming that the effects of fast-fading are averaged out across minor temporal or spatial shifts during the measurement).
These radio samples $r_s(f)$ ($s\in\mathcal{S}$ and $f\in\mathcal{F}$) are reported to a fusion center, which may be, e.g., a base station or a cloud server, depending on the application.

To facilitate mathematical analysis, in this paper, we narrow our focus to a single frequency $f$, considering only spatial samples with RSS $r_s$ for each $s\in\{1,2,\cdots,k\}$ at the corresponding location $y_s\in\mathcal{Y}$.
Moreover, we discretize the continuous space $\mathcal{Y}$ into a $H\times W$ grid (with height $H$ and width $W$).
Hence, each location $y_s$ corresponds to a grid index $(h_s,w_s)$ where $h_s\in\{0,1,\cdots,H-1\}$ and $w_s\in\{0,1,\cdots,W-1\}$.
Define $\mathbf{S}_k$ as an $H\times W$ matrix, where its element at index $(h_s,w_s)$, i.e., $\mathbf{S}_k(h_s,w_s)$, is 0 if there is no sensor located at that grid point; otherwise, it is the RSS value $r_s$ of the sample collected by the sensor $s$ with location $(h_s,w_s)$.  

To ground our findings in reality, we take environmental obstacles, specifically buildings, into consideration. 
These structures can occupy any grid point within the region $\mathcal{Y}$, often covering multiple adjacent points.
We consider the occupied grid points to have no RSS; however, they do affect RSS in surrounding areas by reflecting radio waves back into the environment and by blocking or significantly attenuating radio waves passing through them.
The layout of buildings in this paper is represented by a matrix $\mathbf{B}$, where its element at index $(h_s,w_s)$, i.e., $\mathbf{B}(h_s,w_s)$, is $0$ if there is no building at the point of the grid $(h_s,w_s)$; otherwise, it is $1$.

\begin{figure}[]
	\centering
	\includegraphics[width=\linewidth]{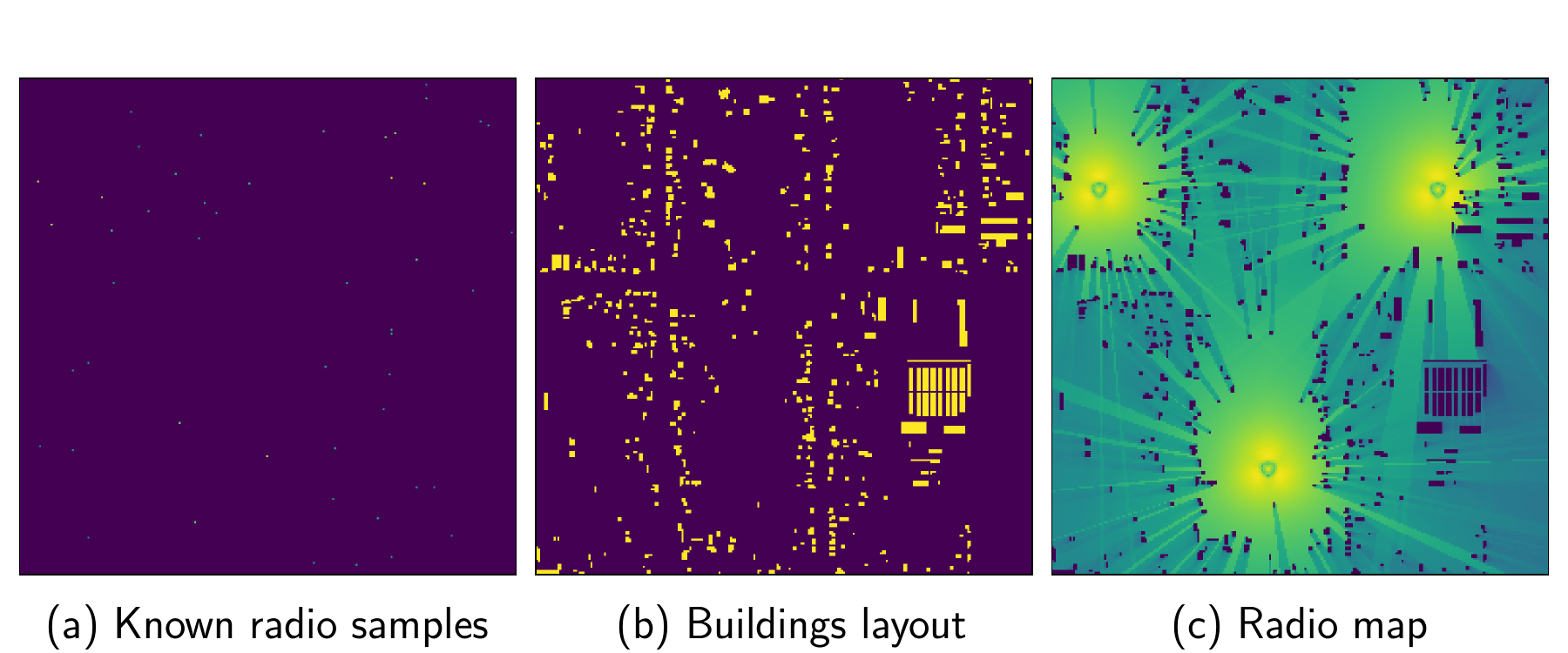}
	\caption{An illustration of the radio map estimation problem.}
	\label{inputAndTarget}
\end{figure}

 {The problem of radio map estimation requires the fusion center to use $\mathbf{S}_k$ and $\mathbf{B}$ to infer (estimate) the RSS at every grid point within the region $\mathcal{Y}$.}
Specifically, as shown in Fig.~\ref{inputAndTarget}, the problem of radio map estimation requires the fusion center to use the information of Fig.~\ref{inputAndTarget}a ($\mathbf{S}_k$, which gives RSS values of collected radio samples) and Fig.~\ref{inputAndTarget}b ($\mathbf{B}$, which is the layout of buildings) to infer Fig.~\ref{inputAndTarget}c (RSS values at each location in the region of interest).

\section{WiFi-Diffusion: Basic Ideas}\label{scheme}

\begin{figure*}[!t]
	\centering
	\includegraphics[width=0.8\linewidth]{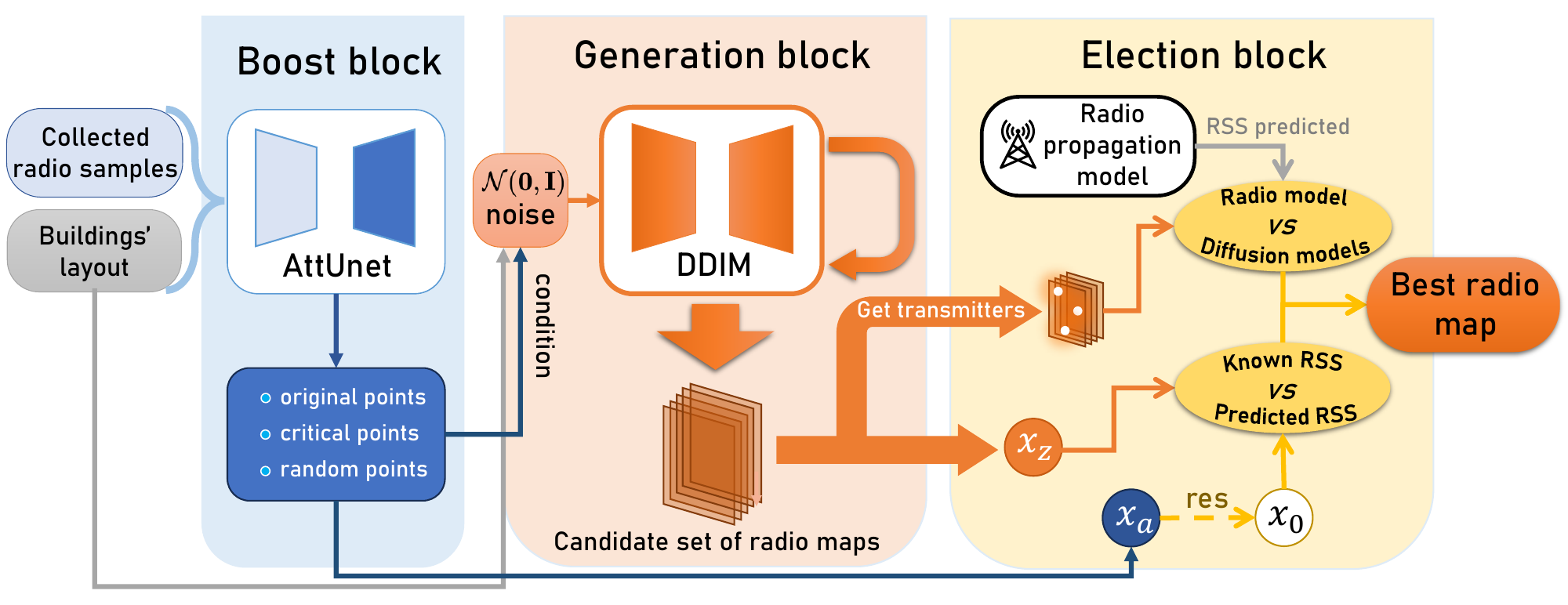}
	\caption{WiFi-Diffusion -- an architecture overview.}
	\label{overview}
\end{figure*}

The problem of fine-grained radio map estimation is extremely challenging because the number of samples collected by the sensors is extremely small compared to the high resolution of radio maps to be estimated in practice.
It is almost impossible for traditional approaches to construct a radio map when only a tiny fraction, e.g., $0.1\%$, of the points in the map are known.
To overcome this challenge, we propose WiFi-Diffusion -— an algorithm that leverages diffusion-based generative techniques to construct fine-grained, high-quality radio maps at ultra-low sampling rates.

As shown in Fig.~\ref{overview}, WiFi-Diffusion consists of three blocks: a boost block, a generation block, and an election block.
In the following, we briefly describe these three blocks.
The objective of the boost block is to optimize and enhance the subsequent generation block.
It would be impossible to discuss the boost block without first gaining a clear understanding of the generation block. 
So we first describe the generation block.

\smallskip

\noindent {\bf Block 2: Generation block\/} \ \ \
In this block, WiFi-Diffusion uses the diffusion-based generative model Denoising Diffusion Implicit Model (DDIM) to produce a candidate set of fine-grained radio maps from the limited known radio samples.
Generative models have an ``out-of-the-box thinking" ability and can make diverse predictions (despite that the majority of which may not be accurate) from extremely limited inputs.
For the problem of fine-grained radio map estimation, we use DDIM to generate multiple fine-grained radio maps (64 in simulations) from only a few (less than $0.1\%$ in simulations) radio samples.
Although most of these generated maps may deviate significantly from the true radio environment, we expect that there are a few high-quality maps within the set.

In the following, we briefly describe the other two blocks of WiFi-Diffusion.

\smallskip

\noindent {\bf Block 1: Boost block\/} \ \ \
Although generative models like DDIM exhibit creativity, the extremely limited number of radio samples often results in a candidate set of maps that are predominantly or entirely of low quality.
Therefore, it is important to design a boost block preceding the generation block, improving the quality of the maps in the candidate set.
Specifically, in the boost block, WiFi-Diffusion leverages prior information, including the layout of buildings, principles of radio propagation, and collected radio samples, as inputs to a UNet model with self-attention.
The UNet model identifies ``critical points" within the geographical region $\mathcal{Y}$ from the perspective of constructing radio maps and conveys the information of these points to DDIM for use in the subsequent generation block.

\smallskip

\noindent {\bf Block 3: Election block\/} \ \ \
Even with the boost block, many maps in the candidate set generated by DDIM are of low quality.
Therefore, it is necessary to design an election block following the generation block to find the best map in the candidate set.
Specifically, in the election block,  WiFi-Diffusion employs the mathematical radio propagation model as a guide to identify the map that most closely adheres to the propagation model in the candidate set.
WiFi-Diffusion selects it as the best map and considers it as the final output.

With the above-mentioned three blocks, WiFi-Diffusion is able to construct fine-grained, high-quality radio maps when the sampling rate is extremely low.
In the following section, we describe the details of the three blocks of WiFi-Diffusion.

\section{WiFi-Diffusion: Design Details}\label{scheme-detail}

WiFi-Diffusion comprises three blocks: a boost block, a generation block, and an election block.
In this section, we describe the design details of each WiFi-Diffusion block.

\subsection{Boost block}

The objective of this block is to improve the quality of fine-grained radio maps produced by DDIM in the subsequent generation block.
In this block, we design AttUnet, a UNet model with self-attention, to pinpoint ``critical points" within the region $\mathcal{Y}$.
AttUnet predicts RSS values at these critical points based on prior information, including the layout of buildings, the law of radio propagation, and RSS values of known radio samples.
These predictions are then provided to DDIM.
Although DDIM, as a diffusion model, is adept at focusing on important features, it lacks a deep understanding of the underlying physical laws.
By incorporating AttUnet's predictions that are aware of physical laws, DDIM's performance can be improved.
Furthermore, AttUnet enhances DDIM by providing it with RSS information of additional radio samples (i.e., with RSS values predicted for the identified critical points). 
Using both identified critical points and limited collected samples, DDIM can generate fine-grained radio maps of higher quality compared to those maps generated solely from the limited collected samples. 

It should be clear that DDIM can more easily predict the propagation of high-power radios in an environment without buildings. 
In contrast, predicting the propagation of low-power radios in a complex environment with buildings poses a greater challenge to DDIM.
We consider the complex environment with buildings as the area containing points that are critical to DDIM.

\begin{figure}[]
	\centering
	\includegraphics[width=\linewidth]{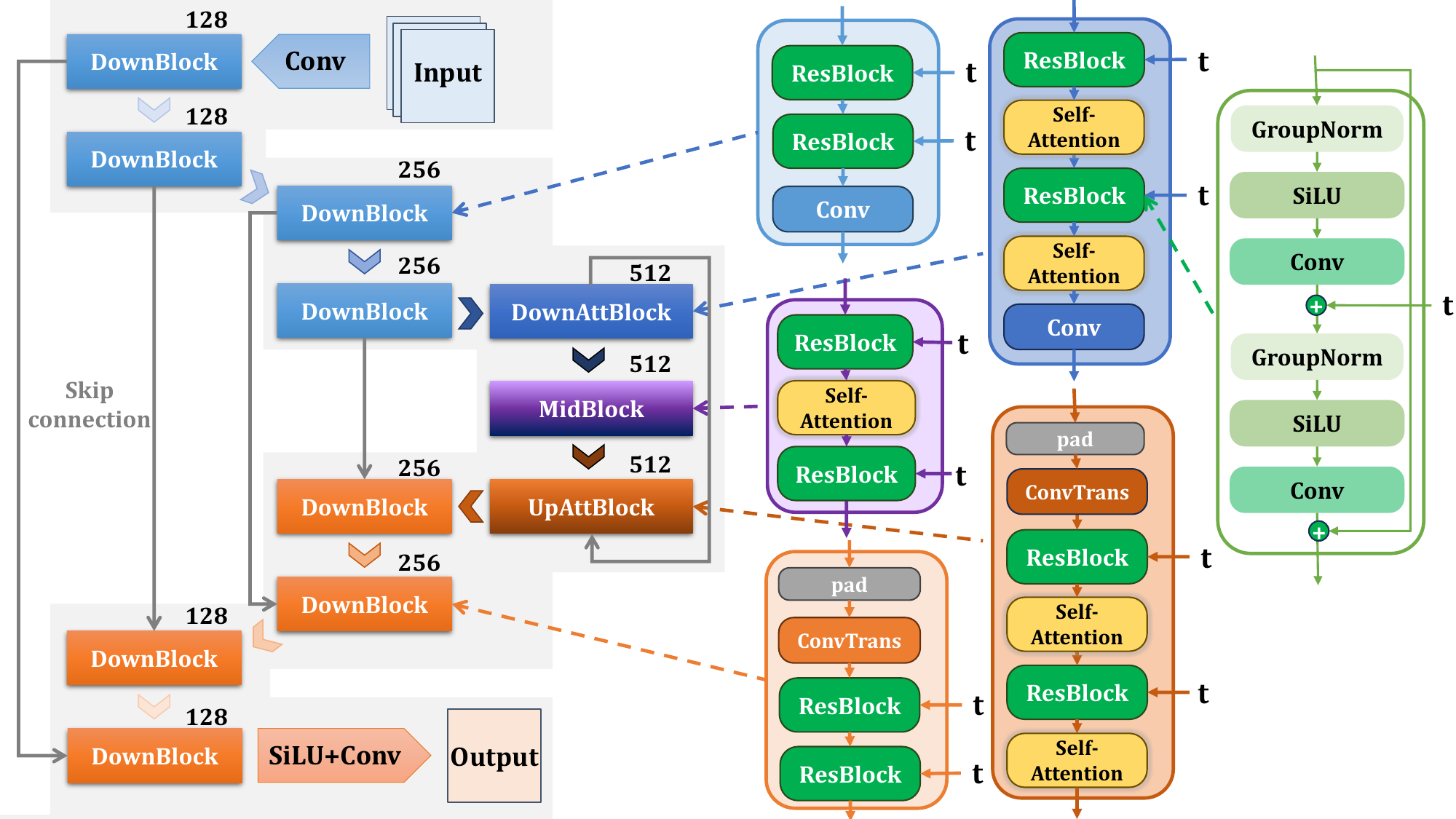}
	\caption{Architecture of AttUnet used in the boost block in WiFi-Diffusion.}
	\label{network}
\end{figure}

\begin{table*}[]
    \centering
    \renewcommand{\arraystretch}{1.3}
    \caption{Parameters of AttUnet.}
    \begin{tabular}{c|ccccccccccccc}
    \hline\hline
        Layer number & Input & 1 & 2 & 3 & 4 & 5 & 6 & 7 & 8 & 9 & 10 & 11 & Output  \\ \hline
        Input Channel & 3 & 128 & 128 & 256 & 256 & 512 & 512 & 512 & 512 & 256 & 256 & 128 & 128  \\ 
        Output Channel & 128 & 128 & 256 & 256 & 512 & 512 & 512 & 512 & 256 & 256 & 128 & 128 & 1  \\ 
        Input Resolution & 256 & 256 & 128 & 64 & 32 & 16 & 8 & 8 & 16 & 32 & 64 & 128 & 256  \\ 
        Output Resolution & 256 & 128 & 64 & 32 & 16 & 8 & 8 & 16 & 32 & 64 & 128 & 256 & 256  \\ 
        With Attention &  &  & &&&$\checkmark$&$\checkmark$&$\checkmark$&&&&&  \\ \hline\hline
    \end{tabular}
    \label{netP}
\end{table*}

The architecture of AttUnet is illustrated in Fig. \ref{network}, and the parameters of AttUnet are detailed in Table \ref{netP}.
Specifically, AttUnet first generates a rough estimate of the fine-grained radio map.
Then, we consider critical points to be located in areas where there is a localized, abrupt change in the RSS or a sudden variation in the rate of change in the RSS on the map. 
These phenomena are typically attributed to intricate interactions between electromagnetic waves and buildings, as well as other waves. 
To identify these critical points, we shift a local window in various directions to analyze the area within the window. 
If this shifting results in substantial changes in the intensity of RSS, we consider the corresponding point to be critical. 

Mathematically, the change of the map $\mathbf x_a$ generated by AttUnet in appearance of the window $w(x,y)$ given the shift $[u,v]$ is
\begin{equation}
    E(u,v)=\sum_{x,y}w(x,y)[\mathbf x_a(x+u,y+v)-\mathbf x_a(x,y)]^2,
\end{equation}
where $\mathbf x_a(x,y)$ is the RSS of the point located at $(x,y)$ in the map $\mathbf x_a$, and $w(x,y)$ is a Gaussian window function which could effectively remove the noise in $\mathbf x_a$ and is suitable for smoothing $\mathbf x_a$ with continuity due to its characteristics of high center weights and low edge weights.
Local quadratic approximation of $E(u,v)$ in the neighborhood of $(0,0)$ is given by the second-order Taylor expansion:
\begin{equation}
\begin{aligned}
    E(u,v)\approx E(0,0)+\begin{bmatrix}
  u&v
\end{bmatrix}\begin{bmatrix}
  E_u'(0,0)\\E_v'(0,0)
\end{bmatrix}+\\\frac{1}{2}\begin{bmatrix}
  u&v
\end{bmatrix}\begin{bmatrix}
  E_{uu}''(0,0)& E_{uv}''(0,0)\\
  E_{uv}''(0,0)& E_{vv}''(0,0)
\end{bmatrix}\begin{bmatrix}
  u\\v
\end{bmatrix},
\end{aligned}
\end{equation}
which can be simplified as 
\begin{equation}
    E(u,v)\approx \begin{bmatrix}
  u&v
\end{bmatrix}\mathbf M\begin{bmatrix}
  u\\v
\end{bmatrix},\end{equation}
where
\begin{equation}
\mathbf M :=\sum_{x,y}w(x,y)\begin{bmatrix}
  ((\mathbf x_a)'_{x})^2& (\mathbf x_a)'_{x}(\mathbf x_a)'_{y}\\
 (\mathbf x_a)'_{x}(\mathbf x_a)'_{y} &((\mathbf x_a)'_{y})^2
\end{bmatrix}.
\end{equation}
According to the method of diagonalization of quadratic forms, we have:
\begin{equation}
\mathbf M =\mathbf R^{-1}\begin{bmatrix}
  \lambda_1& 0\\
 0 &\lambda_2
\end{bmatrix}\mathbf R.
\end{equation}

Given a point $(x,y)$, we first calculate $\mathbf M$ and then compute $\lambda_1$ and $\lambda_2$.
If $\lambda_1$ and $\lambda_2$ are large, the RSS change of the map $\mathbf x_a$ at this point is abrupt, and we consider this point as a critical point.

\begin{figure}[]
	\centering
	\includegraphics[width=\linewidth]{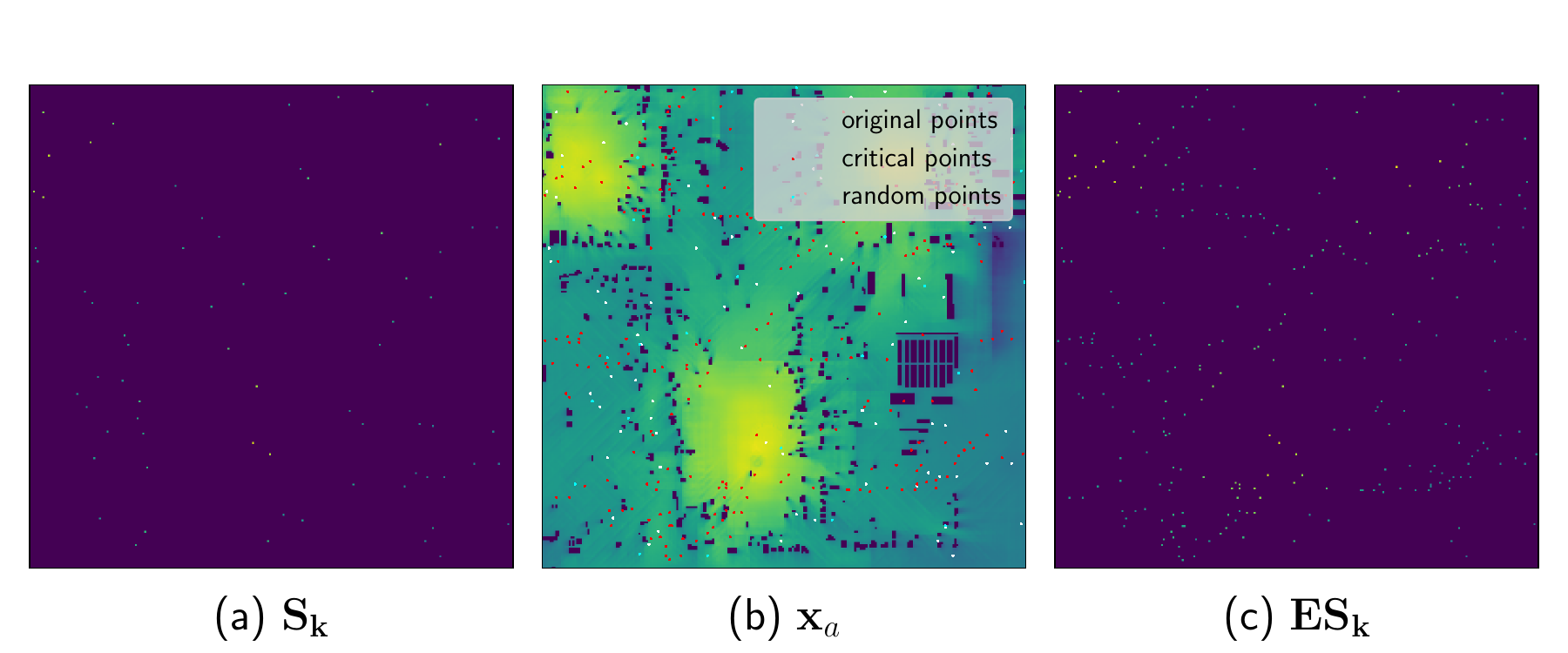}
	\caption{Different kinds of points that are the outputs of the boost block in WiFi-Diffusion.}
	\label{PointEnhance}
\end{figure}

Overall, in the boost block, we first use AttUnet to generate a preliminary fine-grained radio map from the collected known radio samples and buildings layout, and then identify critical points in the map from the perspective of laws of radio propagation.
At the end of the boost block, we convey the predicted radio samples (identified critical points as well as corresponding predicted RSS values) and the known samples collected by the sensors to DDIM.
To further enhance the performance of DDIM, we also randomly select a specific number of points in the map generated by AttUnet and transmit their predicted RSS values to DDIM.
Similarly to $\mathbf{S}_k$, we define $\mathbf{ES}_k$ as a $H\times W$ matrix, where its element at the index $(h_s,w_s)$, i.e., $\mathbf{ES}_k(h_s,w_s)$, represents the RSS value at that grid point if it is a critical point, a randomly selected point, or a point where a radio sample has been collected; otherwise, it is 0.
Fig.~\ref{PointEnhance}a gives an example of $\mathbf{S}_k$ and Fig.~\ref{PointEnhance}c gives an example of $\mathbf{ES}_k$.
Fig.~\ref{PointEnhance}b gives an example of a radio map, where there are original points corresponding to known radio samples, critical points identified in the boost block, and randomly selected points in the map.
$\mathbf{ES}_k$ is the output of the boost block and is the input of the subsequent generation block.

\subsection{Generation block}

The objective of this block is to generate a diverse set of fine-grained radio map candidates.
We employ DDIM to achieve this goal.

We consider the task of fine-grained radio map estimation as a generative problem, in which a fine-grained radio map is regarded as a sample drawn from a $(H\times W)$-dimensional distribution. 
By learning this distribution from available radio map datasets, we can effectively tackle the challenge posed by ultra-low sampling rate and produce fine-grained, high-quality radio maps even when the number of known radio samples in the map to be estimated is extremely small.

We assume that $\mathbf{x}_0$, which is the fine-grained radio map to be estimated, follows a certain conditional probability distribution $\mathcal{X}_c$ given $\mathbf{B}$ and $\mathbf{ES}_k$. 
We use the Denoising Diffusion Probabilistic Model (DDPM)~\cite{DDPM} to learn $\mathcal{X}_c$.
The key idea of DDPM involves transforming $\mathcal{X}_c$ into a standard normal distribution $\mathcal{N}(\mathbf 0,\mathbf I)$, and then training a neural network to model the reverse process, thereby establishing a mapping from $\mathcal{N}(\mathbf 0,\mathbf I)$ back to $\mathcal{X}_c$.
With DDPM, we can get sample values of $\mathcal{X}_c$ by sampling in $\mathcal{N}(\mathbf 0,\mathbf I)$.

Specifically, DDPM consists of two processes: a forward process (diffusion process) and a reverse process (inverse diffusion process). 

\smallskip

\noindent {\bf Forward Process\/} \ \ \
It is a Markov chain. We manually create transition kernels $q(x_t|x_{t-1})$ to incrementally transform data distributions into tractable prior distributions. A typical design of a transition kernel is a Gaussian perturbation:
\begin{equation} q(\mathbf{x}_t\mid\mathbf{x}_{t-1}) = \mathcal{N}(\mathbf{x}_t; \sqrt{1-\beta_t} \mathbf{x}_{t-1}, \beta_t \mathbf{I}),\end{equation}
where $\beta_t$ is the hyperparameter chosen before model training.
Using the chain rule and the Markov property of probability, we denote the joint distribution of $\mathbf{x}_1, \mathbf{x}_2 \dots \mathbf{x}_t$ conditional on $\mathbf{x}_0$ as:
\begin{equation} q(\mathbf{x}_1, \ldots, \mathbf{x}_t\mid\mathbf{x}_0) = \prod_{i=1}^{t} q(\mathbf{x}_i\mid\mathbf{x}_{i-1}). \end{equation}

Based on the reparameterisation trick, we have:
\begin{equation} \left\{\begin{aligned}\mathbf{x}_1&= \sqrt{1-\beta_1}\mathbf{x}_{0}+ \beta_1 \epsilon\\\mathbf{x}_2&= \sqrt{1-\beta_2}\mathbf{x}_{1}+ \beta_2 \epsilon\\&\vdots\\\mathbf{x}_t&= \sqrt{1-\beta_t}\mathbf{x}_{t-1}+ \beta_t \epsilon\end{aligned}\right. \end{equation}
where $\epsilon\sim \mathcal N(\mathbf{0},\mathbf{I})$.
Let $\alpha_t := 1 - \beta_t,\bar{\alpha}_t := \prod_{s=0}^{t} \alpha_s$.
We can marginalize the above joint distribution to give:
\begin{equation} \mathbf{x}_t =\sqrt{\bar{\alpha}_t} \mathbf{x}_0 + \sqrt{1-\bar{\alpha}_t} \epsilon,\end{equation}\begin{equation}q(\mathbf{x}_t\mid\mathbf{x}_0) = \mathcal{N}(\mathbf{x}_t; \sqrt{\bar{\alpha}_t} \mathbf{x}_0, (1-\bar{\alpha}_t) \mathbf{I}). \end{equation}

Considering that
\begin{equation} \lim_{t\rightarrow \infty}\bar{\alpha}_t=\lim_{t\rightarrow \infty} \prod_{s=0}^{t} \alpha_s=\lim_{t\rightarrow \infty} \prod_{s=0}^{t} (1 - \beta_s)=0, \end{equation}
we have
\begin{equation}\lim_{t\rightarrow \infty}\mathbf x_t\sim \mathcal N(\mathbf{0},\mathbf{I})\end{equation}

\smallskip

\noindent {\bf Reverse Process\/} \ \ \
As we can transform $\mathcal{X}_c$ into $\mathcal N(\mathbf{0},\mathbf{I})$ by adding noise, we can likewise map $\mathcal N(\mathbf{0},\mathbf{I})$ to $\mathcal{X}_c$ by removing noise, which is the reverse process of DDPM.

In the reverse process, the true distribution $q(\mathbf x_{t-1}\mid\mathbf x_t)$ is not available. 
However, after adding the conditional posterior probability $\mathbf x_{0}$, it can be obtained by Bayes' Theorem and by the properties of Markov Chains:
\begin{equation}\begin{aligned}q(\mathbf x_{t-1}\mid\mathbf x_t,\mathbf x_0)&=\frac{q(\mathbf x_{t-1},\mathbf x_t,\mathbf x_0)}{q(\mathbf x_t,\mathbf x_0)}\\&=q(\mathbf x_{t}\mid\mathbf x_{t-1})\frac{q(\mathbf x_{t-1}\mid\mathbf x_0)}{q(\mathbf x_t\mid\mathbf x_0)}\\&=\mathcal{N}(\mathbf x_{t-1}; \tilde \mu_t(\mathbf x_t, \mathbf x_0), \tilde\beta_t \mathbf I).\end{aligned}\end{equation}
Thus, the reverse process is solvable if the diffusion is based on a normal distribution.

The inverse Markov chain can be represented by the prior distribution $p(\mathbf x_t) = \mathcal{N}(\mathbf x_t; \mathbf{0}, \mathbf{I})$ and the learnable transfer kernel $p_\theta(\mathbf{x}_{t-1}\mid\mathbf{x}_t)$ (use $p_\theta(\mathbf{x}_{t-1}\mid\mathbf{x}_t)$ to denote $q(\mathbf x_{t-1}\mid\mathbf x_t,\mathbf x_0)$). The learnable transfer kernel is of the form:
\begin{equation} p_\theta(\mathbf x_{t-1}\mid\mathbf x_t) = \mathcal{N}(\mathbf x_{t-1}; \mu_{\theta}(\mathbf x_t, t), \Sigma_{\theta}(\mathbf x_t, t)), \end{equation}
where $\theta$ denotes the model parameters, $\mu_{\theta}(\mathbf x_t, t)$ can be solved by the deep neural network, which has the same architecture as AttUnet, and $\Sigma_{\theta}(\mathbf x_t, t)$is a time-dependent parameter.

Define condition $\{\mathbf{B},\mathbf{ES}_k\}$ as $c$, we have:
\begin{equation} p_\theta(\mathbf x_{t-1}\mid\mathbf x_t,c) = \mathcal{N}(\mathbf x_{t-1}; \mu_{\theta}(\mathbf x_t, t), \Sigma_{\theta}(\mathbf x_t, t)). \end{equation}

During training, we train the 
denoise model with
\begin{equation}\nabla_{\theta}\Vert \epsilon-\epsilon_\theta(\sqrt{\bar{\alpha}_t}x_0+\sqrt{1-\bar{\alpha}_t}\epsilon,t,c) \Vert^2,\end{equation}
where $\epsilon\sim \mathcal N(\mathbf{0},\mathbf{I})$ is the noise, and $\epsilon_\theta$ is the denoise model.

\begin{figure}[]
	\centering
    \subfloat[DDPM-based training.\label{DDPMtrain}]{
		\includegraphics[width=0.48\linewidth]{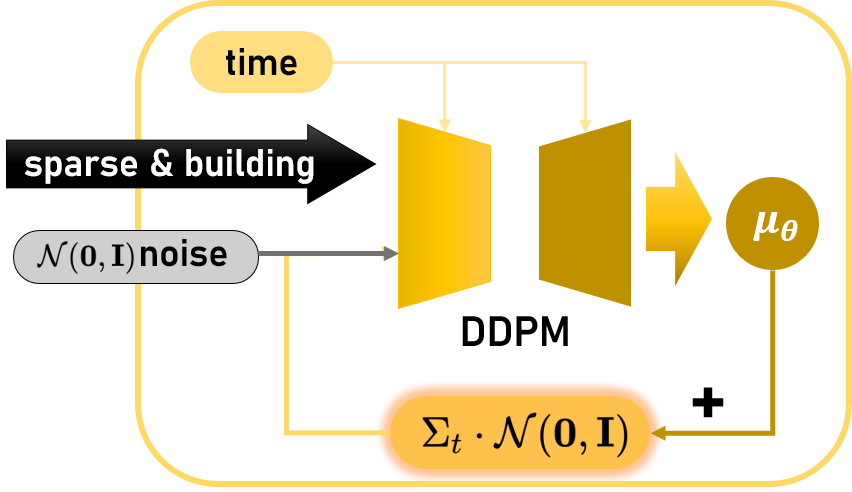}}
\subfloat[DDIM-based inference.\label{DDIMtest}]{
		\includegraphics[width=0.48\linewidth]{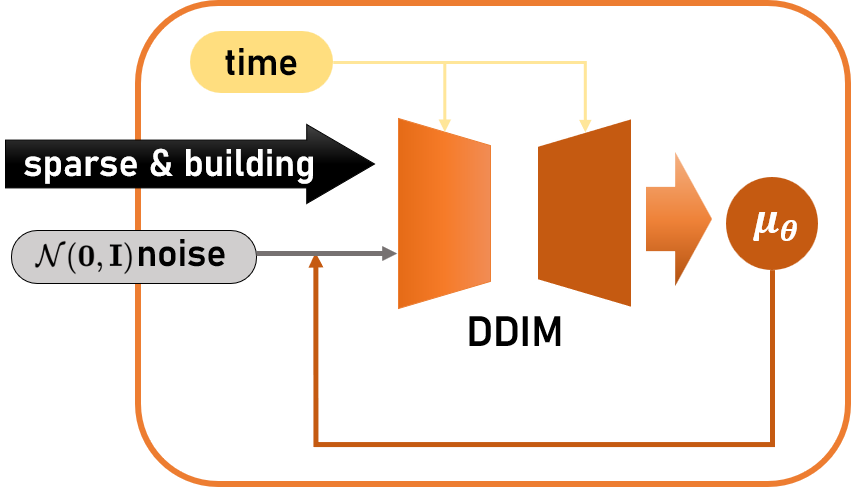}}
	\caption{Training and inference of the generation block in WiFi-Diffusion.}
	\label{DDPMtrainDDIMtest}
\end{figure}

During inference, our aim is to generate a diverse set of results (fine-grained radio maps) from different noises sampled in $\mathcal{N}(\mathbf 0,\mathbf I)$.
A significant limitation of DDPM is the substantial number of iterations required to infer and generate high-quality results.
To address this issue, we use DDIM~\cite{DDIM} for inference.
DDIM markedly reduces the inference time while preserving the high quality of the generated results. 
Furthermore, DDIM can be seamlessly applied after model training, allowing for a straightforward conversion of a DDPM model to a DDIM model without the need for retraining.

DDIM infers $\mathbf x_{t-1}$ from $\mathbf x_t$ as follows
\begin{equation}
    \begin{aligned}\mathbf{x}_{t-1} &= \sqrt{\alpha_{t-1}}\Big(\underbrace{\frac{\mathbf{x}_t-\sqrt{1-\alpha_{t}}\mathbf{\epsilon}_\theta(\mathbf{x}_t, t)}{\sqrt{\alpha_{t}}}}_{\text{predicted}\ \mathbf{x}_0}\Big) \\&+ \underbrace{\sqrt{1 - \alpha_{t-1} - \sigma_t^2} \cdot \mathbf{\epsilon}_\theta(\mathbf{x}_t, t)}_{\text{direction pointing to }\ \mathbf{x}_t} + \underbrace{\sigma_t\epsilon_t}_{\text {random noise}},
    \end{aligned}
\end{equation}
where $\sigma_t^2 = \eta \cdot\sqrt{(1-\alpha_{t-1})/(1-\alpha_{t})}\sqrt{(1-\alpha_{t}/\alpha_{t-1})}$.
If $\eta=1$, the DDIM method is the same as DDPM. 
If $\eta=0$, the DDIM method solves the ordinary differential equations.

The training and inference of the generation block are illustrated in Fig.~\ref{DDPMtrainDDIMtest}.
At the end of the generation block, we get a diverse set of fine-grained radio map candidates generated by DDIM from $\mathbf{B}$ and $\mathbf{ES}_k$. 

\subsection{Election block}
The objective of this block is to identify the best map in the diverse set of fine-grained radio map candidates generated by the generation block.
We leverage the radio propagation model as a guide to identify the map that adheres most closely to the propagation model in the set and consider it the best map.

\begin{figure}[]
	\centering
	\includegraphics[width=2in]{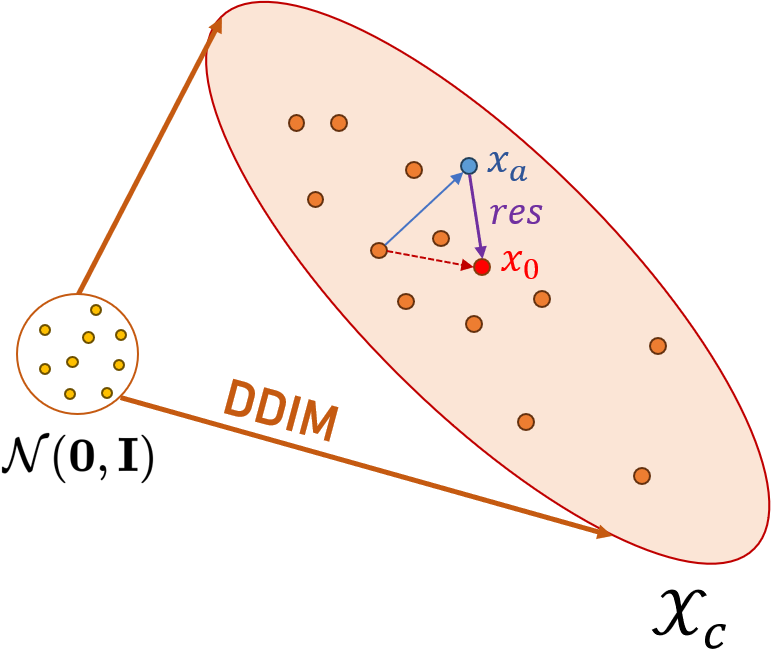}
	\caption{An illustration of the election block in WiFi-Diffusion.}
	\label{WidelySelect}
\end{figure}

As discussed in the previous block and illustrated by Fig.~\ref{WidelySelect}, the diffusion model DDIM can build a mapping from the noisy distribution $\mathcal{N}(\mathbf 0,\mathbf I)$ to the conditional distribution $\mathcal X_c$ under condition $c$:
\begin{equation}
    \mathbf x_{zi}=\text{DDIM}(\mathbf z_i,c),
\end{equation}
where $\mathbf x_{zi}\in \mathcal{X}_c$ and $\mathbf z_i\in \mathcal{N}(\mathbf{0},\mathbf{I})$ with $i\in \{1,2,\cdots,m\}$ and $m$ being the number of map candidates generated by DDIM.

The problem of radio map estimation requires the result $\mathbf x_{zi}$ to be close to the true radio map $\mathbf x_0$.
Therefore, we generate $\{\mathbf x_{z1},\mathbf x_{z2},\cdots,\mathbf x_{zm} \}$ from $\{\mathbf z_1,\mathbf z_2,\cdots,\mathbf z_m \}$, and select $\mathbf x_z\in\{\mathbf x_{z1},\mathbf x_{z2},\cdots,\mathbf x_{zm} \}$ that is closest to $\mathbf x_0$ as output.
Considering that the true map $\mathbf x_0$ is unknown, intuitively, we plan to use the Mean-Square Error (MSE) loss between the RSS values of known radio samples and the RSS values predicted by DDIM to approximate the distance between $\mathbf x_{zi}$ and $\mathbf x_0$, i.e., \begin{equation}\text{MSE}(\mathbf x_{zi},\mathbf x_0 )\sim \text{spMSE}(\mathbf x_{zi},\mathbf{S}_k ) = \text{MSE}(\mathbf x_{zi}\cdot \mathbf{mask}_k,\mathbf{S}_k ),\end{equation}
where $\mathbf{mask}_k$ denotes a mask matrix consisting entirely of zeros, except for the element located at index $(h_s,w_s)$ which corresponds to a non-zero element in $\mathbf{S}_k$.
From the above equation, we have
\begin{equation}
\mathbf x_z=\mathop{\mathrm{argmin}}\limits_{\mathbf x_{zi}}\{\text{spMSE}(\mathbf{S}_k,\mathbf x_{zi})\}.
\end{equation}

\begin{figure}[]
	\centering
	\includegraphics[width=\linewidth]{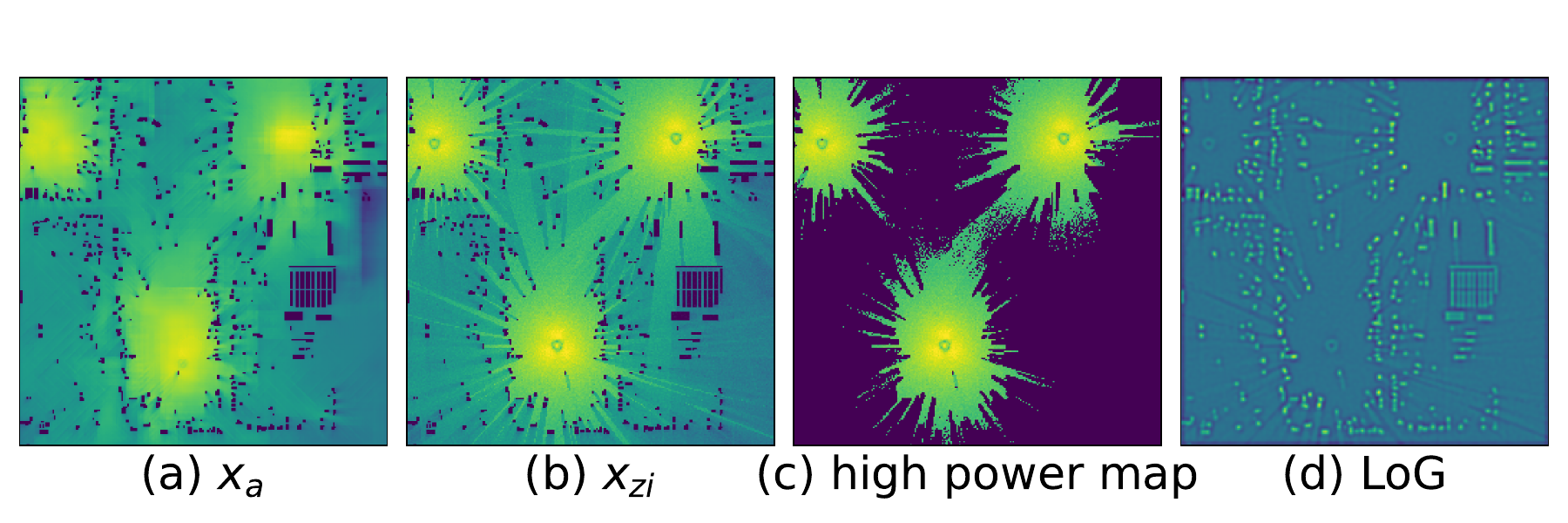}
	\caption{Information used by the election block to learn the correction factor $res$ in WiFi-Diffusion.}
	\label{ResNetInput}
\end{figure}

However, we note that $\text{spMSE}(\mathbf x_{zi},\mathbf{S}_k )$ does not approximate $\text{MSE}(\mathbf x_{zi},\mathbf x_0 )$ well in practice, mainly due to the ultra-low sampling rate (because almost all elements in $\mathbf{S}_k$ are zero).
Recall that in the boost block, AttUnet generates a radio map $\mathbf x_a$. 
Here we use $\mathbf x_a$ to help select the best map.
We introduce a correction factor $res$ as follows
\begin{equation}
\text{MSE}(\mathbf x_0,\mathbf x_{zi})=\text{MSE}(\mathbf x_a,\mathbf x_{zi})+res,
\end{equation}
where the value of $res$ can be derived from different methods of analysis.
ResNet~\cite{ResNet} has been widely used in the past few years in feature extraction applications.
In WiFi-Diffusion, we use it to predict $res$. 
We do this by providing $\mathbf x_a$, $\mathbf x_{zi}$, the Laplacian of Gaussain (LoG) operator processing information (edge detection), and high power maps of $x_{zi}$ (an illustrative example is shown in Fig. \ref{ResNetInput}) to ResNet to learn the correction factor $res$.

In addition to the MSE, we also incorporate a radio propagation model as a guide to select the best map during the election block.
It is important to note that obtaining an exact radio propagation model for the radio map to be estimated is almost impossible, particularly in complex environments.
Instead of focusing on intricate areas (such as low-power zones with buildings) where radio propagation is difficult to predict, we focus on simpler areas (such as high-power zones without buildings) where radio propagation is more straightforward.
Using the radio propagation model to calculate the RSS values of the points in these simpler areas, we can compare these calculated values with those estimated by the radio map given by DDIM.
We then consider the radio map whose estimated RSS values closely align with those calculated by the propagation model as a high-quality radio map.

Specifically, during the election block, we consider areas pertaining to high-power zones located in close proximity to transmitters as simpler areas. 
Given that the vicinity of the transmitter closely approximates the ideal conditions for radio propagation, the RSS distribution within this area is expected to adhere to the Friis Transmission Formula~\cite{Friis}:
\begin{equation}
    \frac{P_r}{P_t}~=~\frac{A_rA_t}{d^2\lambda^2},
\end{equation}
where $P_t$ is the power fed into the transmitting antenna at its input terminals, $P_r$ is the power available at the output terminals of the receiving antenna, $A_r$ is the effective area of the receiving antenna, $A_t$ is the effective area of the transmitting antenna, $d$ is the distance between antennas, and $\lambda$ is the wavelength.

The Friis Transmission Formula pertains to the scenario involving two single-port antennas and is predicated on several key assumptions, including polarization matching, a sufficiently large distance $d$ (satisfying the far-field condition), and the reciprocity of the transmitting antenna. 
In our election block, we employ a generalized version of the Friis Transmission Formula that is applicable to multiport antenna arrays~\cite{FriisM}:
\begin{equation}
    G_{\text{U}} = \frac{P_{r}}{P_{t}} \approx \frac{\mathbf{x}^{\text{H}} \mathbf{y}_{\text{oi}}^{\text{H}} \left( \mathbf{y}_{\text{oo}} + \mathbf{y}_{\text{oo}}^{\text{H}} \right)^{-1} \mathbf{y}_{\text{oi}} \mathbf{x}}{\mathbf{x}^{\text{H}} \left( \mathbf{y}_{\text{ii}} + \mathbf{y}_{\text{ii}}^{\text{H}} \right) \mathbf{x}},
    \label{FM}
\end{equation}
where $\mathbf{y}_{\text{ii}}$ is the self-admittance of the input port, $\mathbf{y}_{\text{oi}}$ is the mutual admittance from the input port to the output port,
and $\mathbf{y}_{\text{oo}}$ is the self-admittance of the output port. 
The inner admittance matrix $\mathbf{y}_{\text{si}}$ and the open circuit voltage $\mathbf{v}_{si} = \mathbf{v}_i + \mathbf{y}_{si}^{-1} \mathbf{i}_i$ are connected at the input port, and $\mathbf{x} = \left( \mathbf{y}_{ii} + \mathbf{y}_{si} \right)^{-1} \mathbf{y}_{si} \mathbf{v}_{si}$. 

In the free-space far-field case, the propagation of radios follows Equation~\eqref{FM} and the proportionality $\mathbf{y}_{\text{oi}}\propto \lambda/d$
\begin{equation}
    \begin{aligned}
G_{U\text{MAX}} &= \mathcal{G} _{A \text{MAX}} \left( \frac{\lambda}{4\pi d} \right)^2\\
G_{U\text{MIN}} &= \mathcal{G}_{A \text{MIN}} \left( \frac{\lambda}{4\pi d} \right)^2,
\end{aligned}
\end{equation}
where $\mathcal{G} _{A \text{MAX}}$ and $\mathcal{G}_{A \text{MIN}}$ are dimensionless functions of the relative orientations of the transmitting and receiving arrays.

\begin{figure}[]
	\centering
	\includegraphics[width=\linewidth]{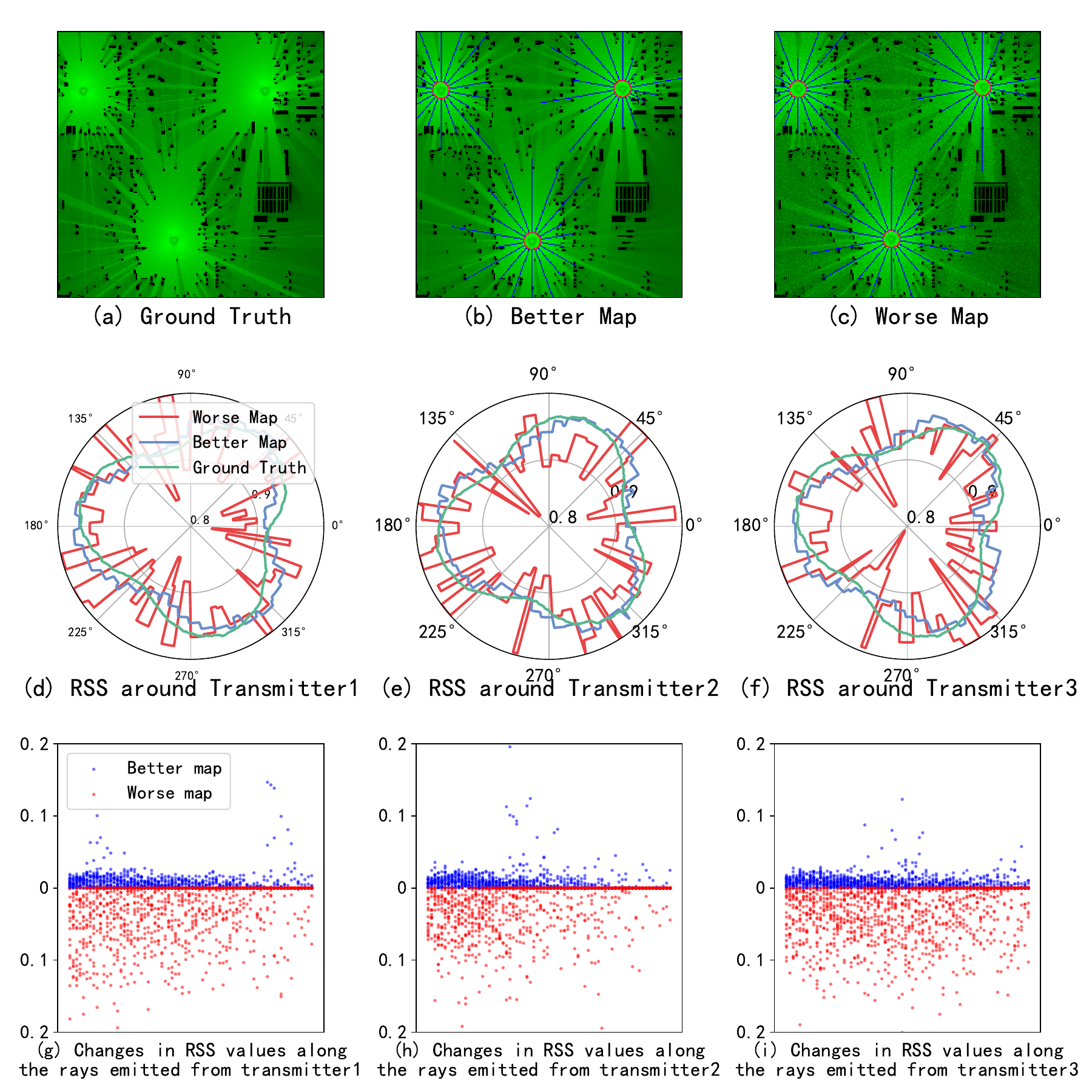}
	\caption{An example illustrating how the election block in WiFi-Diffusion uses laws of radio propagation as a guide to select high-quality radio maps.}
	\label{phy}
\end{figure}

Although the information of transmitters is not input of our studied radio map estimation problem, at the beginning of the election block, we already get a candidate set of fine-grained radio maps from DDIM. 
It is easy to learn the information about the transmitters from each map in the candidate set.
Using the aforementioned principles of radio propagation in conjunction with the transmitter information we have learned, we can predict the RSS distribution, denoted as $\text{RSS}_{phy}(O,d,\theta)$, in the vicinity of the transmitters using polar coordinates. 
Here, $O$ represents the centers of transmitters, and the polar axis aligns with the x-axis of the original data. 

To find the best map in the candidate set of maps generated by DDIM, we focus on the circular regions centered around the transmitters and compute the corresponding RSS values.
We then compare these calculated RSS values with those at the same locations in each candidate map. 
The quality of a map candidate is determined by the degree of similarity, which we quantify using MSE. 
Furthermore, we emit probe rays from the transmitters and examine the variations in RSS as the rays propagate until they intersect with buildings. 
The credibility of our results is evaluated by assessing whether the observed RSS behavior conforms to the smooth attenuation pattern predicted by the physical laws of radio propagation.

Fig.~\ref{phy} gives a simulated example (the specific settings of simulations will be presented later in Section V) to illustrate how the election block uses laws of radio propagation to find the best radio map from the set of map candidates.
Fig.~\ref{phy}a is the true (ground truth) radio map with three transmitters.
Both Fig.~\ref{phy}b and Fig.~\ref{phy}c are estimated radio maps, with Fig.~\ref{phy}b better than Fig.~\ref{phy}c.
In Fig.~\ref{phy}d, we compare the RSS values within a circular region centered around the first transmitter on different maps.
It is evident that the RSS values obtained from the better map (blue results) closely match those of the true map (green results), while the RSS values from the worse map (red results) deviate significantly.
Similar patterns are observed in Fig.~\ref{phy}e and Fig.~\ref{phy}f, which correspond to the circular regions centered around the second and third transmitters, respectively.
Fig.~\ref{phy}g displays the changes in RSS values along the rays emitted from the first transmitter.
It is clear that the changes in RSS values in the better map (blue results) are much closer to zero compared to those in the worse map (red results).
Similar observations can be found in Fig.~\ref{phy}h and Fig.~\ref{phy}i, respectively.

Overall, we define an indicator $\text{rate}_{phy}$ as follows
\begin{equation}
\begin{aligned}
    \text{rate}_{phy}=&\frac{1}{n}\sum_n^i\alpha\cdot\int_0^{2\pi}(\text{RSS}_{phy}(O_i,d,\theta)-\mathbf x_{zi}(O_i,d,\theta))^2\\&+(1-\alpha)\cdot\sum_l^{\text{num. of rays}}\left\lfloor\frac{\partial\mathbf x_{zi}}{\partial l}/\sigma\right\rfloor,
\end{aligned}
\end{equation}
where $n$ is the number of transmitters, $\alpha$ is a hyperparameter, and $\sigma$ is a threshold of volatility intensity.
In the election block, we calculate the MSE between $\mathbf x_a$ and $\mathbf{x}_{zi}$, as well as $\text{rate}_{phy}$ to find the best map $\mathbf x_z$.
Specifically, we define
\begin{equation}\label{eqn:election}
\mathbf x_z = \left\{ \begin{array}{r}
	\mathop{\mathrm{argmin}}\limits_{\mathbf x_{zi}}\{\lambda\cdot \text{spMSE}(\mathbf{S}_k,\mathbf x_{zi})+\hat{\lambda}\cdot \text{rate}_{phy}\},\\\text{if}~~k>100\\
	\mathop{\mathrm{argmin}}\limits_{\mathbf x_{zi}}\{\lambda\cdot \text{MSE}(\mathbf x_a,\mathbf x_{zi})+res+\hat{\lambda}\cdot \text{rate}_{phy}\},\\\text{if}~~k\le 100
\end{array}
\right.
\end{equation}
where $\hat{\lambda}=(1-\lambda)$.

\subsection{Summary}
In summary, WiFi-Diffusion is a three-block diffusion model-based algorithm for fine-grained radio map estimation.
The three blocks of WiFi-Diffusion are the boost block, the generation block, and the election block.
In the boost block, WiFi-Diffusion utilizes prior information, e.g., layout of buildings, principles of radio propagation, and collected radio samples, to enhance the qualities of fine-grained radio map candidates generated by the diffusion model.
In the generation block, WiFi-Diffusion leverages the creative power of the diffusion-based generative model to produce a diverse set of fine-grained radio map candidates.
In the election block, WiFi-Diffusion uses the mathematical radio propagation model as a guide to find the best fine-grained map from the diverse candidates generated by the diffusion model.
In the following section, we use extensive simulations to verify that WiFi-Diffusion is able to generate fine-grained, high-quality radio maps at ultra-low sampling rates, and show that it significantly outperforms SOTA.

\section{Evaluation}\label{Experimental}

In this section, we evaluate WiFi-Diffusion.
We first use a case study to evaluate its performance and demonstrate its capability of constructing fine-grained, high-quality radio maps at ultra-low sampling rates.
Then we use extensive simulations to compare it against various baseline algorithms, highlighting that it significantly outperforms SOTA at ultra-low sampling rates.
Next, we conduct ablation studies to analyze the individual contributions of the boost block and the election block to its overall performance.
In addition, we perform robustness studies to evaluate its performance under multiple extreme scenarios for radio map estimation, such as varying numbers of transmitters on one map, transmitters located outside the map, and extremely limited number of collected radio samples (as few as 3 samples).
 {Finally, we present its model size and training cost, and use simulations to show that it maintains significant performance advantages over SOTA, even when SOTA is scaled up to comparable or greater complexity and computational expense.}

\subsection{Simulation Settings}

We use the open-source dataset \emph{BART-Lab radio maps}\footnote{https://github.com/BRATLab-UCD/Radiomap-Data} from~\cite{DataSet,dataset2} for the evaluation. 
This dataset is obtained by the simulation software \emph{Altair Feko}~\cite{altair_winprop_2021} from \emph{WinProp} and consists of 2000 radio maps.
The number of transmitters on one map varies from $1$ to $3$.
There are 2000 different layouts of buildings in the dataset, extracted from \emph{OpenStreetMap}~\cite{openstreetmap} in the United States.
The heights of the buildings are set as $10$ m.
This dataset consists of radio maps simulated in 5 different frequency bands: 1750 MHz, 2750 MHz, 3750 MHz, 4750 MHz, and 5750 MHz. 
For our evaluation, we specifically use the 5750 MHz (WiFi 5G) data.
Considering that the resolutions of the original maps in the dataset are different, to ensure consistency, we randomly select regions with a resolution of $256 \times 256$ (i.e., $H=W=256$) from each map.
In total, we have 1776 different radio maps, each of which has a resolution of $256 \times 256$.
For WiFi-Diffusion, we use 1300 maps for training, 376 maps for testing, and 100 maps for inference (evaluation), respectively.

We consider the number of samples collected by sensors, i.e., $k$, to be ultra-small: 
\begin{equation}
k~\in~\{200, 150, 100, 50, 25, 10\}.\nonumber   
\end{equation}

Since the resolution of each radio map is $256\times 256$, the sampling rate in evaluation is
\begin{equation}
    \frac{k}{H\times W}\in\{0.31\%,0.23\%,0.15\%,0.08\%,0.04\%,0.02\%\}.\nonumber
\end{equation}

We implement WiFi-Diffusion on an NVIDIA RTX 4090 GPU and an AMD EPYC 9654 CPU. 
For WiFi-Diffusion, deep learning codes are built using PyTorch, the number of DDPM inference steps is 1000 ($t=1000$), and the number of DDIM inference steps is 10 ($u=10$).
There are 64 map candidates generated by DDIM in the WiFi-Diffusion generation block for each problem instance ($m=64$).

To evaluate WiFi-Diffusion, we compare it with the following interpolation-based baseline algorithms:
\begin{itemize}
    \item RBF~\cite{rbf}: By creating a network of basis functions around the known data points, it uses a linear combination of these basis functions to approximate the values of unknown points.

    \item Splines~\cite{spline}: It uses piecewise low-degree polynomials to approximate the values of unknown points on the curve or surface between known data points.

    \item Ordinary kriging~\cite{kriging}: It uses a semi-variogram function to describe the spatial autocorrelation and estimates the values of unknown points by weighted average of the observed data of known points.
\end{itemize}

We also compare WiFi-Diffusion with the following deep learning-based baseline algorithms:

\begin{itemize}
    \item RadioUnet~\cite{RadioUNet}: It uses UNet to estimate radio maps.
    UNet is a convolutional neural network (CNN) architecture, characterized by its symmetric U-shaped structure, consisting of an ``encoder" (contracting path) and a ``decoder" (expansive path).

    \item SkipResidualAutoencoder~\cite{autoencoder}: It uses an auto-encoder to build radio maps. The autoencoder is a type of neural network model that is primarily designed to learn effective representations of data.
    The number of hidden layers in this baseline is set to 128.

    \item ResNet~\cite{ResNet}: ResNet34 is also used for evaluation.
    We change the last FC layer to two MLP layers linked with the SiLU activation function and activate the output with a sigmoid function.
    To convert the extracted features to a $256\times 256$ map, the size of the last layer is set to 65536. 
\end{itemize}

To quantify the performance of algorithms, MSE, Root Mean Square Error (RMSE), Normalized Mean Square Error (NMSE), and Peak Signal to Noise Ratio (PSNR) \cite{psnr} are used:
\begin{itemize}
\item MSE: It is the average of the squared difference between the original true values and the predicted values.
\begin{equation}
    \text{MSE}=\frac{1}{H\times W}\sum^{H-1}_{h=0}\sum^{W-1}_{w=0}[I(h,w)-\hat{I}(h,w)]^2,
\end{equation}
where $I(h,w)$ is the original true RSS value at the point $(h,w)$ and $\hat{I}(h,w)$ is the predicted RSS value at that point.
\item RMSE: It is the square root of MSE.
\begin{equation}\text{RMSE}=\sqrt{\text{MSE}}.\end{equation}
\item NMSE: It is the normalized MSE.
\begin{equation}\text{NMSE}=\frac{\displaystyle\sum^{H-1}_{h=0}\sum^{W-1}_{w=0}[I(h,w)-\hat{I}(h,w)]^2}{\displaystyle\sum^{H-1}_{h=0}\sum^{W-1}_{w=0}I(h,w)^2}.\end{equation}
\item PSNR: It estimates the ratio between the maximum RSS and RMSE, and is usually expressed as a logarithmic quantity using the decibel scale.
\begin{equation}\text{PSNR}=20\cdot\log_{10}\left(\frac{\text{MAX}_I}{\text{RMSE}}\right),\end{equation}
where $\text{MAX}_I$ is the maximum RSS value of a point on the radio map. 
\end{itemize}

From the above definitions, it is clear that a small MSE, a small RMSE, a small NMSE, or a large PSNR implies that the constructed radio map is of high quality.

\begin{figure*}[]
	\centering
	\includegraphics[width=\linewidth]{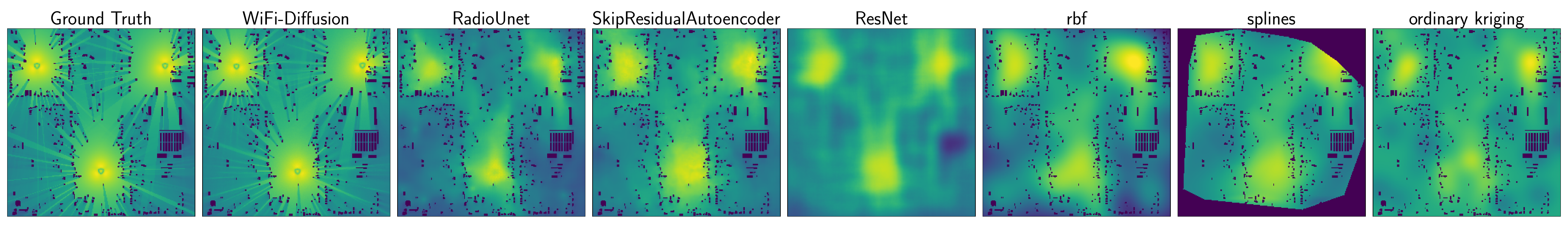}
	\caption{Radio maps constructed by different algorithms in the case study ($k=50$).}
	\label{allVisualResults}
\end{figure*}

\subsection{Simulation Results}

\smallskip

\noindent {\bf Case Study\/} \ \ \
In the case study, we have $k=50$ sensors randomly distributed on a map (the sampling rate is $0.08\%$).  
Fig.~\ref{allVisualResults} shows the radio maps constructed by different algorithms, respectively.
From the figure, it is clear that the radio map constructed by WiFi-Diffusion is very close to the ground truth (true radio map).
It is also clear that the radio map constructed by WiFi-Diffusion is much more accurate than that constructed by any of the 6 baseline algorithms.

\smallskip

\noindent {\bf More Simulations\/} \ \ \
Then we perform more simulations to evaluate WiFi-Diffusion.
Table \ref{mainTable} presents the simulation results comparing WiFi-Diffusion with the baseline algorithms under varying $k$ (i.e., number of samples).
For each $k$, $100$ radio maps are used for evaluation, and the reported metrics (MSE, RMSE, NMSE, and PSNR) represent the average values obtained from these 100 maps.
From the table, we observe that WiFi-Diffusion outperforms all baseline algorithms for every value of $k$, ranging from 10 to 200.
Moreover, the radio maps generated by WiFi-Diffusion exhibit significantly superior quality compared to those produced by the baseline approaches. 
Specifically, at ultra-low sampling rates, the MSE values for WiFi-Diffusion can be up to $10$ times lower than (i.e., at most one-tenth of) those of the baseline methods.
For example,
\begin{itemize}
    \item WiFi-Diffusion vs RBF: The MSE of WiFi-Diffusion is $0.0027$, which is over 15 times lower than $0.0429$ that is the MSE of RBF when $k=25$.

    \item WiFi-Diffusion vs Spline: The MSE of WiFi-Diffusion is $0.0027$, which is over 35 times lower than $0.0948$ that is the MSE of spline when $k=25$.

    \item WiFi-Diffusion vs Ordinary kriging: The MSE of WiFi-Diffusion is $0.0008$, which is over 15 times lower than $0.0126$ which is the MSE of ordinary kriging when $k=100$.

    \item WiFi-Diffusion vs ResNet: The MSE of WiFi-Diffusion is $0.0008$, which is over 10 times lower than $0.0084$ that is the MSE of ResNet when $k=100$.

    \item WiFi-Diffusion vs RadioUnet: The MSE of WiFi-Diffusion is $0.0027$, which is over 15 times lower than $0.0450$ that is the MSE of RadioUnet when $k=25$.

    \item WiFi-Diffusion vs SkipResidualAutoencoder: The MSE of WiFi-Diffusion is $0.0027$, which is over 10 times lower than $0.0239$ that is the MSE of SkipResidualAutoencoder when $k=25$.
\end{itemize}

\begin{table*}
	\centering
	\renewcommand{\arraystretch}{1.3}
	\caption{Simulation results comparing WiFi-Diffusion with baseline algorithms under varying number of radio samples.}
	\label{mainTable}
	\begin{tabular*}{\textwidth}{cp{0.88cm}p{0.88cm}p{0.88cm}p{0.88cm}p{0.88cm}p{0.88cm}p{0.88cm}p{0.88cm}p{0.88cm}p{0.88cm}p{0.88cm}p{0.88cm}}
		\hline\hline
		Number of samples $k$ & \multicolumn{4}{c}{10} & \multicolumn{4}{c}{25}  & \multicolumn{4}{c}{50}     \\
		\cmidrule(r){2-5}  \cmidrule(r){6-9} \cmidrule(r){10-13}\noalign{\smallskip} 
		Metric        & MSE$\downarrow$ & RMSE$\downarrow$ & NMSE$\downarrow$ & PSNR$\uparrow$ & MSE$\downarrow$ & RMSE$\downarrow$ & NMSE$\downarrow$ & PSNR$\uparrow$ & MSE$\downarrow$ & RMSE$\downarrow$ & NMSE$\downarrow$ & PSNR$\uparrow$ \\\hline
		RBF&$0.0535$&$0.2314$&$0.2012$&$12.70$&$0.0429$&$0.2073$&$0.1556$&$13.66$&$0.0361$&$0.1901$&$0.1299$&$14.41$\\
		Spline&$0.1522$&$0.3901$&$0.5646$&$8.175$&$0.0948$&$0.3079$&$0.3450$&$10.23$&$0.0627$&$0.2505$&$0.2340$&$12.02$\\
		Ordinary kriging&$0.0144$&$0.1201$&$0.0530$&$18.40$&$0.0108$&$0.1043$&$0.0402$&$19.62$&$0.0101$&$0.1008$&$0.0385$&$19.92$\\
		ResNet&$0.0142$&$0.1192$&$0.0542$&$18.46$&$0.0104$&$0.1020$&$0.0389$&$19.81$&$0.0092$&$0.0959$&$0.0336$&$20.36$\\
		RadioUnet&$0.1005$&$0.3171$&$0.3696$&$9.975$&$0.0450$&$0.2122$&$0.1649$&$13.46$&$0.0159$&$0.1262$&$0.0590$&$17.97$\\
		Autoencoder&$0.0646$&$0.2542$&$0.2367$&$11.89$&$0.0239$&$0.1546$&$0.0906$&$16.21$&$0.0071$&$0.0843$&$0.0259$&$21.47$\\
		WiFi-Diffusion& $\pmb{0.0079}$& $\pmb{0.0892}$& $\pmb{0.0293}$& $\pmb{20.98}$& $\pmb{0.0027}$& $\pmb{0.0527}$& $\pmb{0.0106}$& $\pmb{25.56}$& $\pmb{0.0013}$& $\pmb{0.0365}$& $\pmb{0.0049}$& $\pmb{28.73}$\\
		\hline\hline Number of samples $k$ &\multicolumn{4}{c}{100}  & \multicolumn{4}{c}{150}   & \multicolumn{4}{c}{200}\\
		\cmidrule(r){2-5}  \cmidrule(r){6-9} \cmidrule(r){10-13}\noalign{\smallskip} 
		Metric        & MSE$\downarrow$ & RMSE$\downarrow$ & NMSE$\downarrow$ & PSNR$\uparrow$ & MSE$\downarrow$ & RMSE$\downarrow$ & NMSE$\downarrow$ & PSNR$\uparrow$ & MSE$\downarrow$ & RMSE$\downarrow$ & NMSE$\downarrow$ & PSNR$\uparrow$ \\\hline RBF&$0.0377$&$0.1944$&$0.1406$&$14.22$&$0.0330$&$0.1817$&$0.1199$&$14.81$&$0.0312$&$0.1767$&$0.1189$&$15.05$\\
		Spline&$0.0405$&$0.2013$&$0.1513$&$13.92$&$0.0313$&$0.1771$&$0.1144$&$15.03$&$0.0271$&$0.1648$&$0.1003$&$15.65$\\
		Ordinary kriging&$0.0126$&$0.1126$&$0.0465$&$18.96$&$0.0194$&$0.1393$&$0.0716$&$17.12$&$0.0283$&$0.1682$&$0.1036$&$15.48$\\
		ResNet&$0.0084$&$0.0921$&$0.0311$&$20.70$&$0.0080$&$0.0895$&$0.0297$&$20.96$&$0.0085$&$0.0922$&$0.0316$&$20.70$\\
		RadioUnet&$0.0029$&$0.0541$&$0.0110$&$25.32$&$0.0013$&$0.0370$&$0.0048$&$28.61$&$0.0011$&$0.0333$&$0.0040$&$29.52$\\
		Autoencoder&$0.0021$&$0.0468$&$0.0080$&$26.58$&$0.0016$&$0.0404$&$0.0061$&$27.86$&$0.0013$&$0.0369$&$0.0050$&$28.65$\\
		WiFi-Diffusion& $\pmb{0.0008}$& $\pmb{0.0287}$& $\pmb{0.0029}$& $\pmb{30.81}$& $\pmb{0.0007}$& $\pmb{0.0265}$& $\pmb{0.0025}$& $\pmb{31.52}$& $\pmb{0.0006}$& $\pmb{0.0252}$& $\pmb{0.0023}$& $\pmb{31.95}$\\
		\hline\hline\end{tabular*}
\end{table*}

In addition, the number of samples needed by WiFi-Diffusion can be at most one-fifth of that needed by each baseline approach to construct comparable radio maps (radio maps with similar MSE).
For example,
\begin{itemize}
    \item WiFi-Diffusion vs RBF: To construct a radio map with MSE no greater than $0.0080$, WiFi-Diffusion needs $10$ samples, while RBF needs over $200$ samples.

    \item WiFi-Diffusion vs Spline: To construct a radio map with MSE no greater than $0.0080$, WiFi-Diffusion needs $10$ samples, while Spline needs over $200$ samples.

    \item WiFi-Diffusion vs Ordinary kriging: To construct a radio map with MSE no greater than $0.0080$, WiFi-Diffusion needs $10$ samples, while ordinary kriging needs over $200$ samples.

    \item WiFi-Diffusion vs ResNet: To construct a radio map with MSE no greater than $0.0080$, WiFi-Diffusion needs $10$ samples, while ResNet needs $150$ samples.

    \item WiFi-Diffusion vs RadioUnet: To construct a radio map with MSE no greater than $0.0080$, WiFi-Diffusion needs $10$ samples, while RadioUnet needs $100$ samples.

    \item WiFi-Diffusion vs SkipResidualAutoencoder: To construct a radio map with MSE no greater than $0.0080$, WiFi-Diffusion needs $10$ samples, while SkipResidualAutoencoder needs $50$ samples.
\end{itemize}

\begin{figure*}[]
	\centering
    \includegraphics[width=\linewidth]{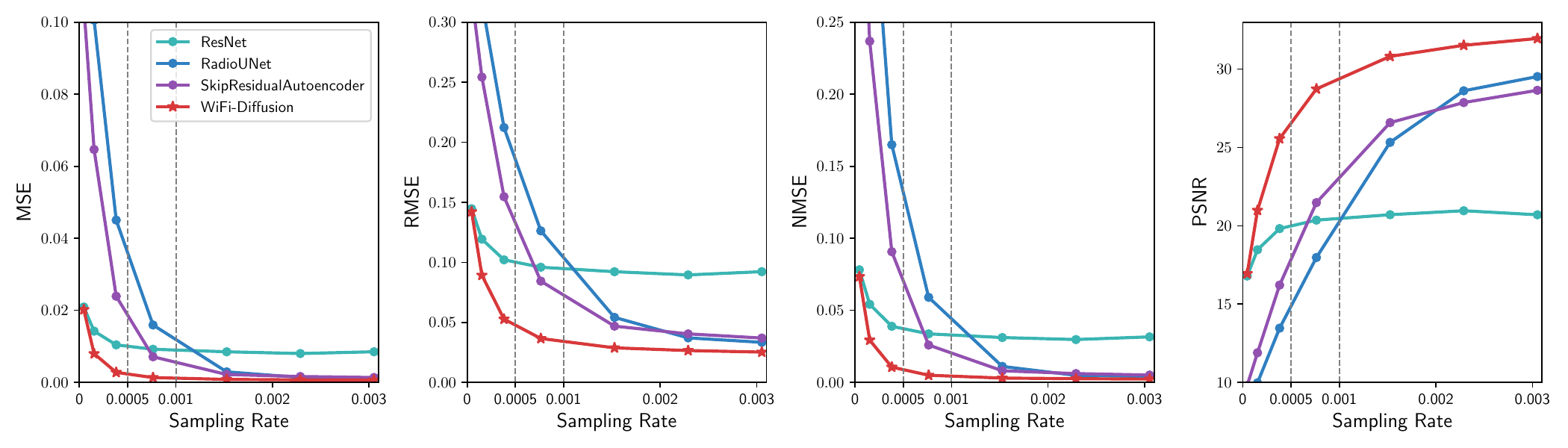}
	\caption{Results comparing WiFi-Diffusion with deep learning-based baselines under varying sampling rates.}
	\label{mainPlot}
\end{figure*}

Fig. \ref{mainPlot} presents curves of simulation results that compare WiFi-Diffusion with the three deep learning-based baseline algorithms under varying $k$.
 {From the figure, we observe that the MSE achieved by WiFi-Diffusion is close to $0$ when $k\ge 50$.
This implies that WiFi-Diffusion generates fine-grained, high-quality radio maps at sampling rates lower than $0.1\%$.
In comparison, the MSE of ResNet is significantly greater than $0$ for any $k$.
The MSE values of RadioUnet and SkipResidualAutoencoder are close to $0$ when $k\ge  150$, but are much greater than $0$ when $k\le 100$.
As a result, none of the three baseline algorithms can generate fine-grained, high-quality radio maps at sampling rates lower than $0.1\%$.}
Similar conclusions can also be drawn for RMSE, NMSE, and PSNR, respectively, based on the results in Fig. \ref{mainPlot}.

\smallskip
\noindent {\bf Ablation Study\/} \ \ \
WiFi-Diffusion consists of three blocks.
In addition to the important diffusion model-based generation block, it has a boost block to optimize the performance of the generation block and an election block to find the best map from the set of maps produced by the generation block.
Now we use ablation studies to evaluate the boost block and the election block, respectively, of WiFi-Diffusion.

\begin{table}[]
	\renewcommand{\arraystretch}{1.3}
	\caption{Ablation studies to evaluate different blocks of WiFi-Diffusion.}
	\label{abTable}
	\centering
	\begin{tabular}{cc|ccc}
		\hline\hline
		\multicolumn{2}{c|}{Setting}   & \multicolumn{3}{c}{MSE under varying number of samples}        \\\hline
		Boost block & Election block & 10      & 25      & 50          \\ \hline
		$\times$ &  $\times$ & $0.03170$ & $0.01980$ & $0.00912$\\
$\checkmark$ & $\times$ & $0.01246$ & $0.00392$ & $0.00199$\\
$\times$ & $\checkmark$  & $0.02359$ & $0.01198$ & $0.00611$\\
$\checkmark$ & $\checkmark$ & $\pmb{0.00796}$ & $\pmb{0.00277}$ & $\pmb{0.00133}$\\
		\hline Boost block & Election block      & 100     & 150     & 200     \\ \hline
	$\times$ & $\times$ & $0.00226$ & $0.00125$ & $0.00111$\\
$\checkmark$ & $\times$ & $0.00139$ & $0.00125$ & $0.00128$\\
$\times$ & $\checkmark$  & $0.00200$ & $0.00089$ & $0.00065$\\
$\checkmark$ & $\checkmark$ & $\pmb{0.00082}$ & $\pmb{0.00070}$ & $\pmb{0.00063}$\\
	\hline\hline\end{tabular}
\end{table}

\begin{table}[]
\renewcommand{\arraystretch}{1.1}
\caption{Simulation results to verify the effectiveness of the election block of WiFi-Diffusion.}
\label{tab:election}
\centering
\begin{tabular}{c|ccc}
\hline\hline
Setting   & \multicolumn{3}{c}{MSE under varying number of samples}        \\\hline
   & 10      & 25      & 50          \\ \hline
 No election block & $0.01246$ & $0.00392$ & $0.00199$\\
\thead{With an election block\\that minimizes~\eqref{eqn:simple-election}}  & $0.00970$ & $0.00292$ & $0.00138$\\
\thead{With our proposed election\\block defined by~\eqref{eqn:election}} & $\pmb{0.00796}$ & $\pmb{0.00277}$ & $\pmb{0.00133}$\\
\hline      & 100     & 150     & 200     \\ \hline
 No election block & $0.00139$ & $0.00125$ & $0.00128$\\
\thead{With an election block\\that minimizes~\eqref{eqn:simple-election}}  & $0.00093$ & $0.00073$ & $0.00068$\\
\thead{With our proposed election\\block defined by~\eqref{eqn:election}} & $\pmb{0.00082}$ & $\pmb{0.00070}$ & $\pmb{0.00063}$\\
\hline\hline\end{tabular}
\end{table}

Table \ref{abTable} gives the simulation results of the ablation studies, which clearly demonstrate the necessity of the boost block and the election block in WiFi-Diffusion.
The performance of WiFi-Diffusion is significantly degraded when either the boost block or the election block is removed, highlighting their critical roles in the overall functionality of WiFi-Diffusion.

\begin{figure*}[]
	\centering
	\includegraphics[width=1\linewidth]{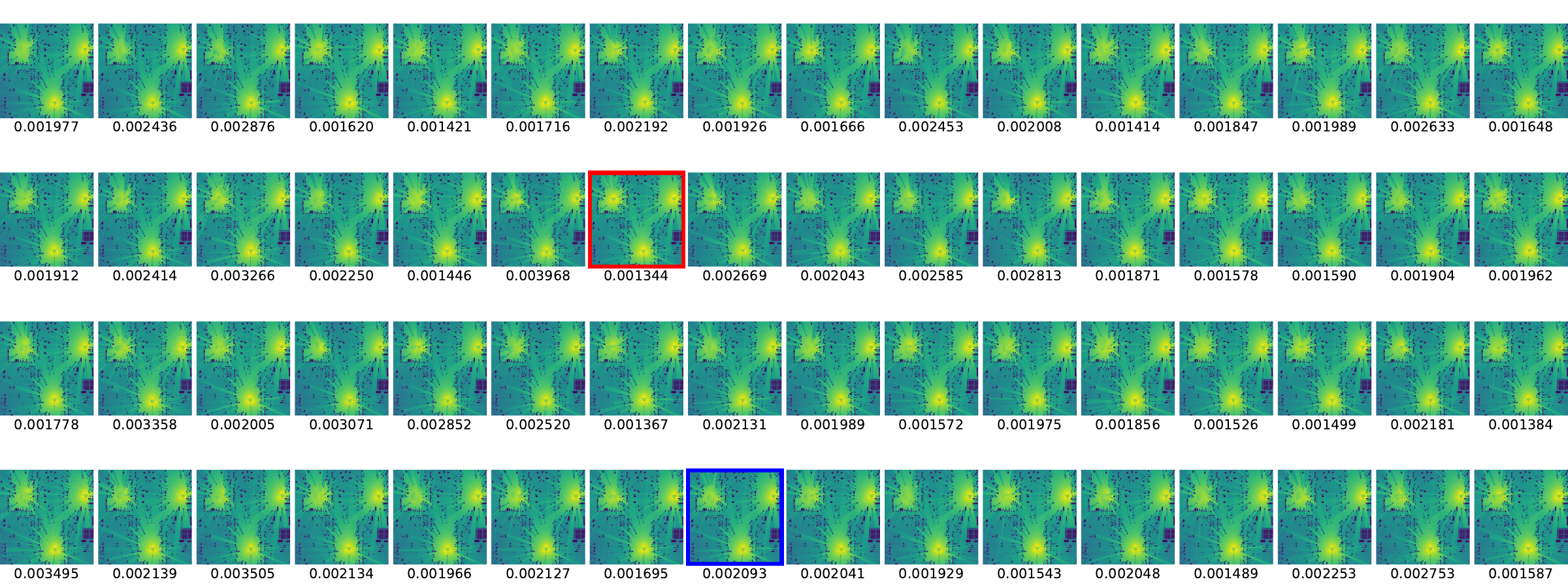}
	\caption{The set of $64$ radio map candidates (as well as the achieved MSE values) generated by the generation block of WiFi-Diffusion in one simulation instance when $k=25$ (the red map is the map identified by the election block and is the best map in the set; the blue map is the map with the lowest $\text{MSE}_r$ defined in~\eqref{eqn:simple-election} and is much worse than the red map).}
	\label{selectPic}
\end{figure*}

 {In the following, we use simulations to further verify the effectiveness of the election block.
Considering that the true radio map (map to be predicted) is unknown during inference, different from our proposed design of the election block, one straightforward design is to identify the candidate map with predicted RSS values closest to the radio samples collected by the sensors.
Specifically, we can calculate $\text{MSE}_r$ defined as follows:
\begin{equation}\label{eqn:simple-election}
    \text{MSE}_r~=~\frac{1}{k}\sum^{k}_{i=1}[I_i-\hat{I}_i]^2,
\end{equation}
where $I_i$ represents the measured RSS value from sensor $i$ and $\hat{I}_i$ denotes the predicted RSS value at the corresponding location in the generated map.
The optimal map is identified as the candidate that minimizes $\text{MSE}_r$.}

 {Fig.~\ref{selectPic} illustrates a representative simulation instance, displaying the candidate set of 64 fine-grained radio maps generated by the generation block of WiFi-Diffusion along with their corresponding MSE values.
It demonstrates a significant variation in map quality, with the best map achieving an MSE of $0.0013$ compared to $0.0040$ for the poorest map, indicating that it is necessary to have an election block in WiFi-Diffusion.
Our proposed election design successfully identifies the optimal map (red map) with the lowest MSE of $0.0013$, while the simple election design based solely on the minimization of $\text{MSE}_r$ selects an inferior candidate (blue map) with an MSE of $0.0021$.}

 {Moreover, we simulate more instances under varying $k$ that is the number of samples, with the results summarized in Table~\ref{tab:election}.
The results consistently show that our election block design outperforms the straightforward $\text{MSE}_r$-based baseline, particularly when $k=10$, $k=25$, and $k=100$.
In general, Fig.~\ref{selectPic} and Table~\ref{tab:election} demonstrate both the necessity of incorporating an election block in WiFi-Diffusion and the superior performance of our proposed design over alternative approaches.}

 {The election block can improve the performance of WiFi-Diffusion by using the law of radio propagation as a guide to select the high-quality radio map from a set of candidate maps produced by the generation block. 
The three-block mechanism of WiFi-Diffusion stems from the inherent strength of generative AI approaches like diffusion models in producing diverse, innovative solutions.
WiFi-Diffusion effectively leverages this generative advantage, which is in contrast to conventional discriminative models such as RadioUnet, SkipResidualAutoencoder, and ResNet.
Although these discriminative approaches offer stable predictions, their nature fundamentally limits their ability to generate diverse output variations.
Consequently, it is difficult to add similar election blocks to them to improve their performance.}

\begin{figure*}[]
	\centering
    \subfloat[There is only 1 transmitter.\label{r-1-1}]{
		\includegraphics[width=0.48\linewidth]{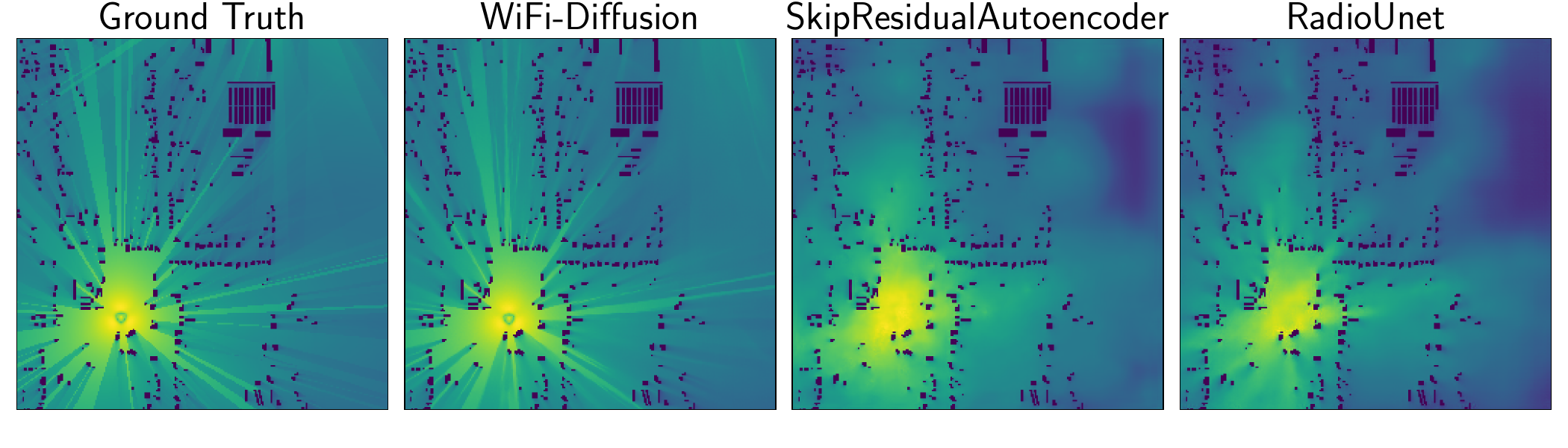}}
	\subfloat[There are 2 transmitters.\label{r-1-2}]{
		\includegraphics[width=0.48\linewidth]{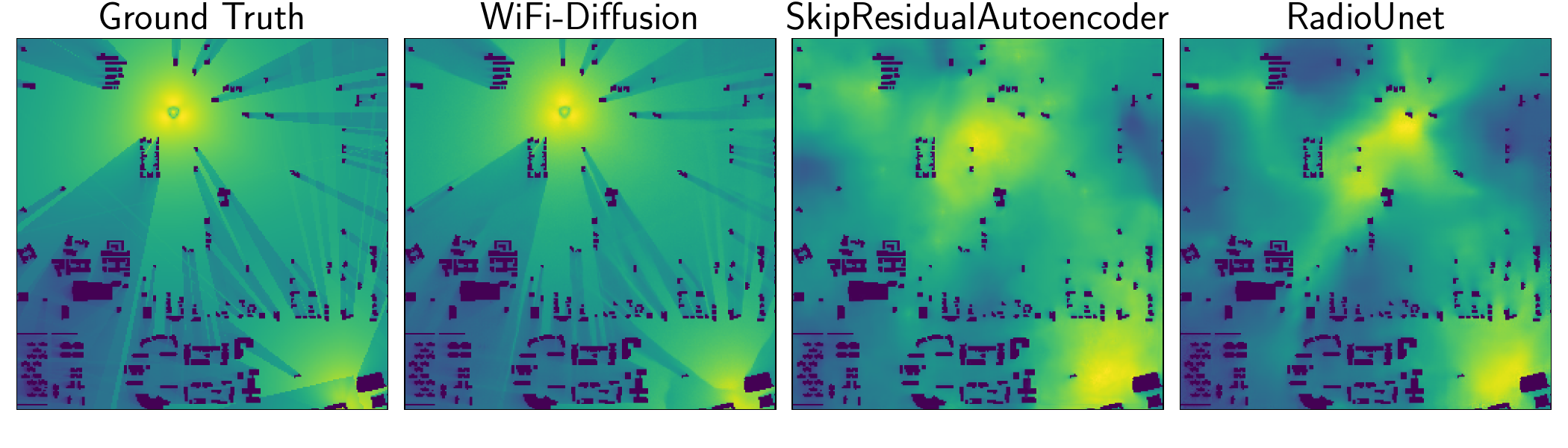}}
	\caption{Radio maps constructed by different algorithms under varying number of transmitters ($k=50$).}
	\label{robustness-1}
\end{figure*}

\begin{figure*}[]
	\centering
    \subfloat[No transmitters are located within the map but 2 outside.\label{r-2-1}]{
		\includegraphics[width=0.48\linewidth]{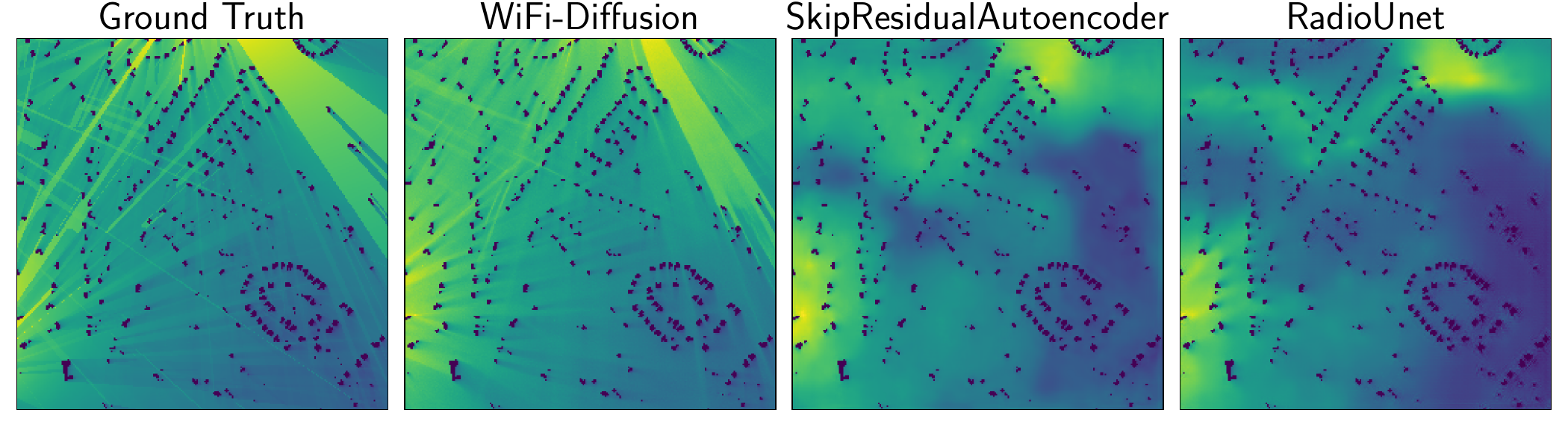}}
	\subfloat[2 transmitters are located within the map and 1 outside.\label{r-2-2}]{
		\includegraphics[width=0.48\linewidth]{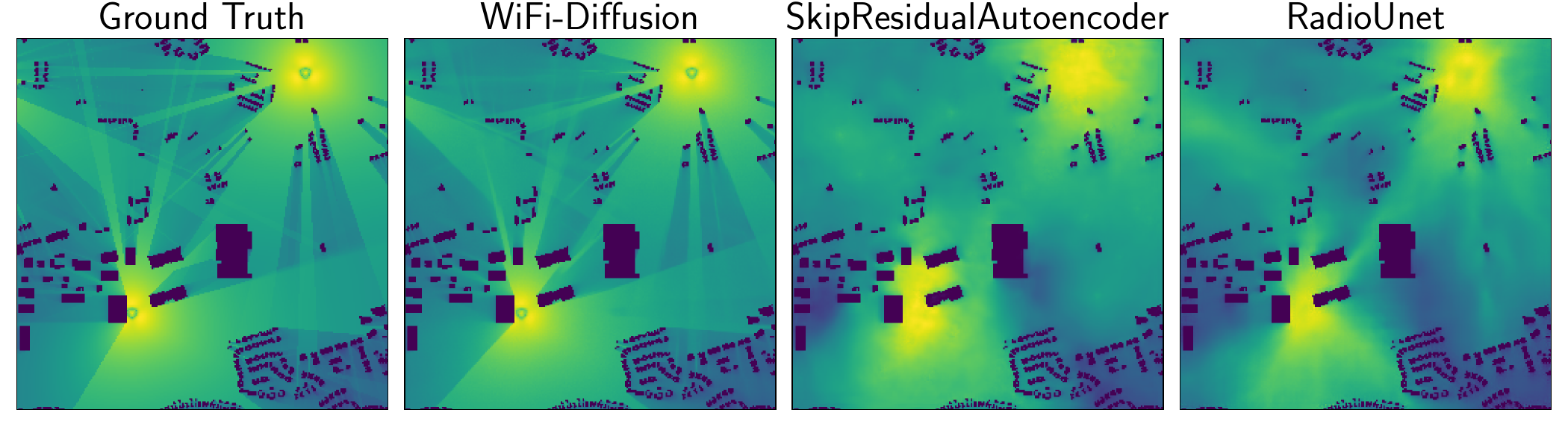}}
	\caption{Radio maps constructed by different algorithms where certain transmitters are located outside the map ($k=50$).}
	\label{robustness-2}
\end{figure*}

\smallskip
\noindent {\bf Robustness Study\/} \ \ \ 
We emphasize that our studied radio map estimation problem does not rely on prior knowledge of transmitter details, such as their quantity or positions.
Here, we are interested in evaluating the performance of WiFi-Diffusion across various transmitter configurations.

Fig. \ref{robustness-1} shows radio maps generated by WiFi-Diffusion and three deep learning-based baseline algorithms, under scenarios with different numbers of transmitters. 
Specifically, Fig. \ref{r-1-1} presents results with a single transmitter, while Fig. \ref{r-1-2} illustrates results with two transmitters, both with $k=50$.
The results demonstrate that WiFi-Diffusion consistently produces fine-grained, high-quality radio maps, regardless of the number of transmitters, and significantly outperforms the baseline algorithms.

Furthermore, Fig. \ref{robustness-2} evaluates the robustness of WiFi-Diffusion when certain transmitters are located outside the map.
In Fig. \ref{r-2-1}, no transmitters are located within the map but two outside, while in Fig. \ref{r-2-2}, two transmitters are located within the map and one outside, again with $k=50$.
Fig. \ref{robustness-2} indicates that WiFi-Diffusion is able to construct fine-grained, high-quality radio maps even when certain transmitters are located outside the map, significantly outperforming baseline algorithms.

\begin{figure}[]
	\centering
	\includegraphics[width=1\linewidth]{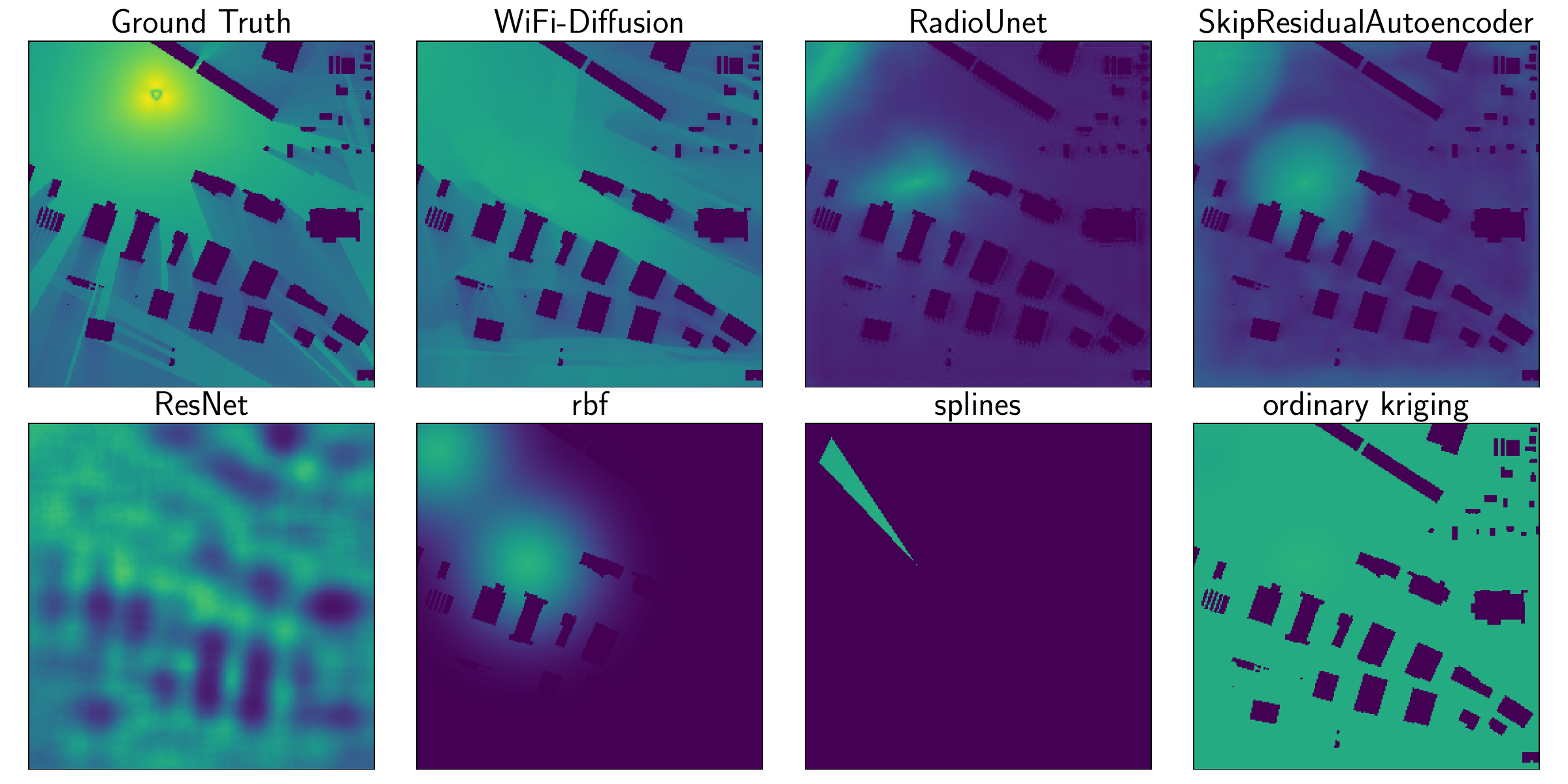}
	\caption{Radio maps constructed by different algorithms when there are only $3$ collected radio samples.}
	\label{robustness0}
\end{figure}

Additionally, we evaluate WiFi-Diffusion in an extremely challenging scenario where the sampling rate is nearly $0$, specifically when $k=3$.
The results of this evaluation are presented in Fig. \ref{robustness0}. 
Despite the daunting task of predicting RSS values for a $256\times 256$ grid using only $3$ known samples, WiFi-Diffusion manages to produce a rough radio map.
In contrast, all six baseline algorithms fail to generate even a basic representation of the radio map. 
This comparison underscores the exceptional robustness of WiFi-Diffusion in estimating radio maps, particularly in the extremely challenging scenario when the sampling rate is nearly $0$.

\begin{table}[]
    \centering
    \caption{Number of parameters, time of inference, and time of training of different algorithms (m: million; s: second; h: hour).}
    \label{table:cost}
    \begin{tabular}{ccccc}
    \hline\hline
        & RadioUnet & Autoencoder & ResNet & WiFi-Diffusion \\ \hline
        \specialcell{Number of\\ parameters} & 13m & 4m & 89m &  122m \\ 
        \specialcell{Time of\\inference} & 0.06s & 0.06s &  0.07s &  7.82s \\ 
        \specialcell{Time of\\training} & 19h & 7h & 11h & 44h \\ 
        \hline\hline
    \end{tabular}
\end{table}

\smallskip
\noindent {\bf Model Size and Training Cost \/} \ \ \ 
 {Due to their inherent complexity, diffusion models typically require more parameters and longer computational times compared to traditional deep learning architectures such as UNet.
As shown in Table~\ref{table:cost}, which compares the model size (number of parameters)\footnote{ {In Table~\ref{table:cost}, the number of parameters of WiFi-Diffusion is near $122$ million, where the number of parameters for the boost block is near $61$ million and that for the generation block is near $61$ million.}}, inference time, and training time across different algorithms, WiFi-Diffusion is much larger and slower than the baseline methods.
A critical question arises:} 

 {\emph{Would baseline methods perform comparably or even outperform WiFi-Diffusion if their parameter count were increased to match or exceed that of WiFi-Diffusion?}}

 {We address this question empirically through comprehensive evaluations.}

 {We enhance each of the three baseline models -- RadioUnet, Autoencoder, and ResNet -- by increasing their depth (number of layers), expanding their width (channel dimensions), and training them with parameter counts and computational time comparable to or exceeding those of WiFi-Diffusion. 
The resulting improved variants are denoted as RadioUnet+, Autoencoder+, and ResNet+, respectively.
As demonstrated in Table~\ref{table:efficiency}, which summarizes the model size (number of parameters), inference time, and training time across all evaluated algorithms, the enhanced baselines (RadioUnet+, Autoencoder+, and ResNet+) exhibit larger model sizes and longer training durations than WiFi-Diffusion.} 

\begin{figure*}[]
	\centering
    \subfloat[$k=25$.]{
		\includegraphics[width=0.4\linewidth]{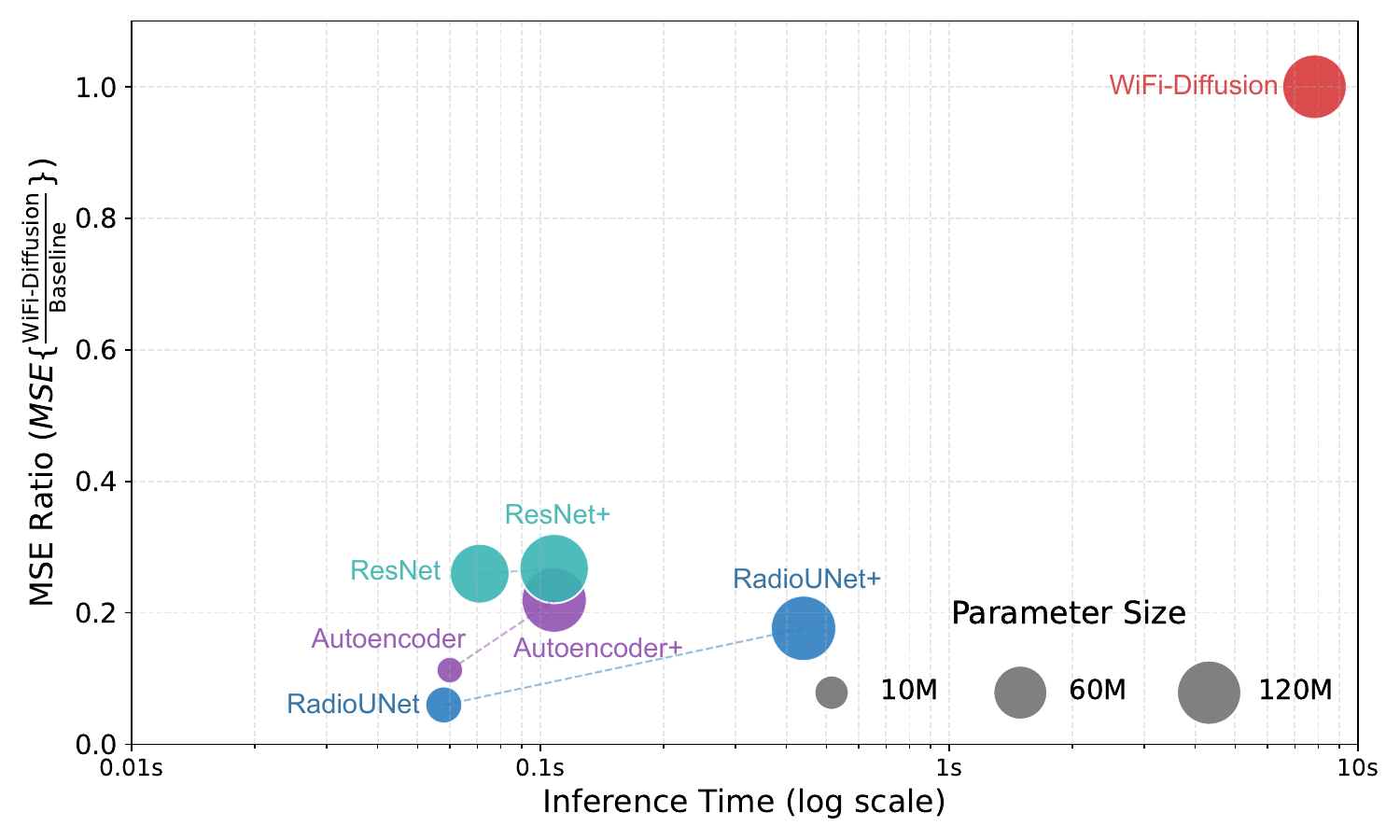}}
	\subfloat[$k=50$.]{
		\includegraphics[width=0.4\linewidth]{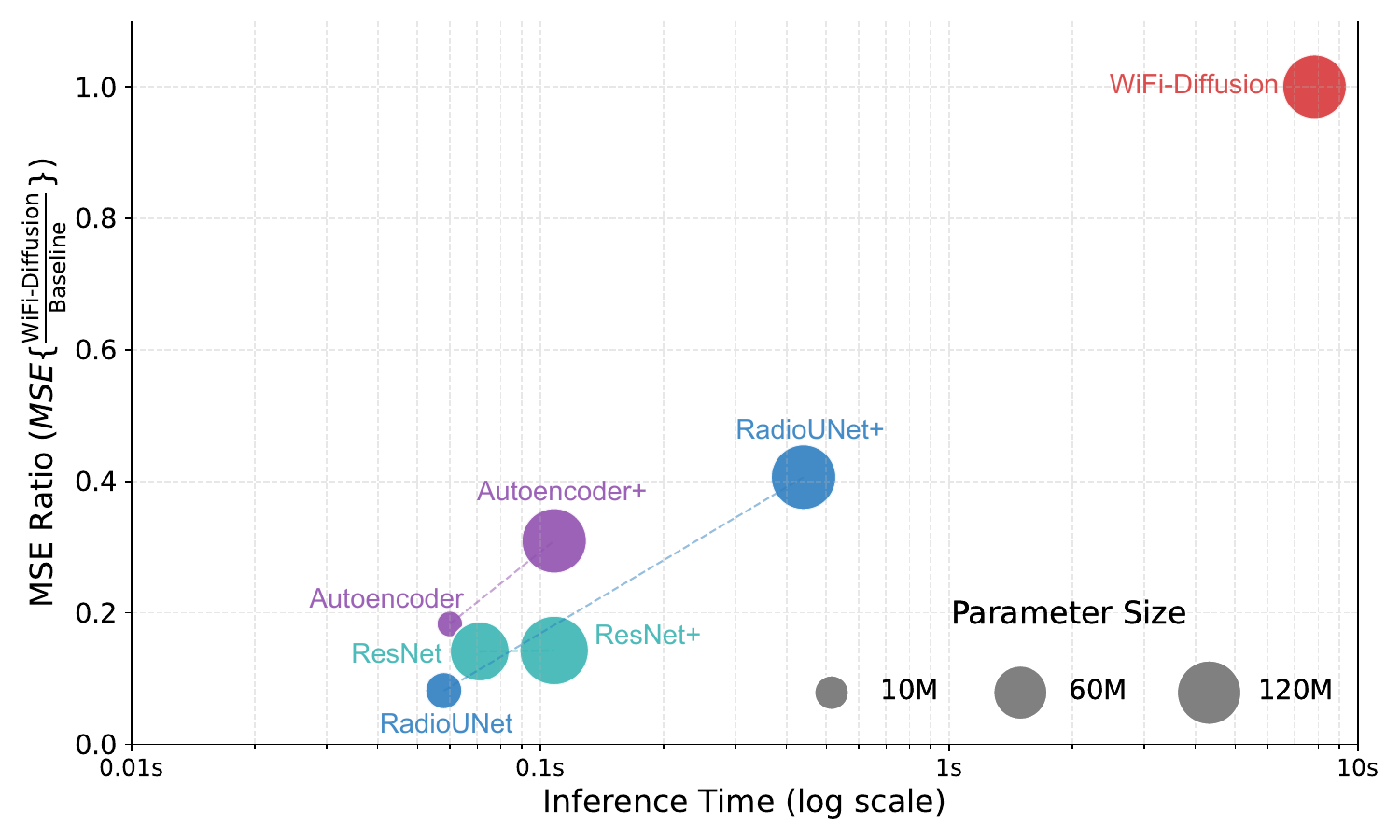}}
	\caption{Simulation results of MSE and inference time achieved by different algorithms.}
	\label{fig:efficiency}
\end{figure*}

 {Fig.~\ref{fig:efficiency} compares the inference times across different algorithms and presents the ratio between the MSE values achieved by WiFi-Diffusion and those achieved by the baseline methods.
Table~\ref{table:efficiency} also shows the results of inference times and MSE values achieved by different algorithms.
From Fig.~\ref{fig:efficiency} and Table~\ref{table:efficiency} we observe that
(i) the radio map constructed by WiFi-Diffusion is significantly better than that constructed by any of the enhanced baselines (RadioUnet+, Autoencoder+, and ResNet+), considering that the MSE achieved by WiFi-Diffusion is $70\%$ lower than that achieved by any of the enhanced baselines when $k=25$ and the MSE achieved by WiFi-Diffusion is $60\%$ lower than that achieved by any of the enhanced baselines when $k=50$;
(ii) While WiFi-Diffusion requires a longer inference time than any of the enhanced baselines, the absolute inference time of WiFi-Diffusion remains practical and acceptable, never exceeding $10$ seconds, enabling real-time radio map generation.
Moreover, it is worth highlighting that manually increasing the inference time for any of the enhanced baseline methods does not yield improvements in the quality of constructed radio maps.
}

\begin{table*}[]
    \centering
    \caption{Number of parameters, time of inference, time of training, and MSE results of different algorithms  (m: million; s: second; h: hour).}
    \label{table:efficiency}
    \renewcommand{\arraystretch}{1.3}
    \begin{tabular}{cccccc}
    \hline\hline
        Method & Number of parameters & Inference time & MSE ($k=25$) & MSE ($k=50$) & Training time \\ \hline
        RadioUnet & 13,274,031 (13m) & 0.06s & 0.0450 &  0.0159 & 19h \\ 
        RadioUnet+ & 144,249,256 (144m) & 0.44s & 0.0153 & 0.0032 & 48h \\ 
        Autoencoder & 3,553,945 (4m) & 0.06s &  0.0239 &  0.0071 & 7h \\ 
        Autoencoder+ & 127,491,225 (127m) & 0.11s & 0.0123 & 0.0042 & 55h \\ 
        ResNet & 88,981,248 (89m) & 0.07s &  0.0104 & 0.0092 & 11h \\ 
        ResNet+ & 161,984,512 (162m) & 0.11s & 0.0101 & 0.0091 & 45h \\ 
        WiFi-Diffusion & 60,823,169 $\times$ 2 (122m) & 7.82s & 0.0027 & 0.0013 & 44h \\ \hline\hline
    \end{tabular}
\end{table*}

\subsection{Summary}
 {In summary, the evaluations given in this section demonstrate that:}

\begin{itemize}
    \item  {WiFi-Diffusion is capable of generating fine-grained, high-quality radio maps when the sampling rate is less than $0.1\%$.}
    
    \item  {WiFi-Diffusion achieves superior accuracy, yielding radio maps with MSE values an order of magnitude lower than those produced by the baseline methods under same ultra-low sampling rates.}
    
    \item  {WiFi-Diffusion achieves comparable accuracy (MSE) to the baseline methods while requiring only $20\%$ of their sampling rate for fine-grained radio map generation.}

    \item  {All three blocks including the boost block, generation block, and election block are essential for WiFi-Diffusion's high-performance on radio map estimation.}

    \item  {Even when the baseline methods are scaled up to match or exceed WiFi-Diffusion's model size and training cost, WiFi-Diffusion maintains significantly superior accuracy, achieving radio maps with MSE values over $60\%$ lower than those produced by the baseline methods.}
\end{itemize}

\section{WiFi-Diffusion: Potential Applications}\label{sec:app}
{WiFi-Diffusion generates fine-grained, high-quality radio maps which present parameters of interest in communication channels, such as the RSS, at every point of a certain geographical region for the WiFi spectrum.
It can improve the performance of many WiFi-based wireless networks, including unmanned aerial vehicle (UAV) network, industrial Internet of Things (IoT), and Internet of Vehicles (IoV).
}

\smallskip
\noindent {\bf {UAV Network} \/} \ \ \
{It serves as the critical communication infrastructure which enables wireless connectivity and information exchange among multiple drones~\cite{UAV-Network}.
Modern UAVs equipped with onboard sensors are capable of collecting and transmitting radio samples.
An edge server can leverage these collected samples through WiFi-Diffusion to construct and disseminate fine-grained, accurate radio maps to individual UAVs in real-time.
These radio maps provide UAVs with predictive capabilities for RSS values in various frequency bands along their designated flight trajectories.
Consequently, each UAV can dynamically optimize its spectrum access strategy by adaptively adjusting transmission parameters according to spatial and temporal variations.
In this case, WiFi-Diffusion can be used to improve key performance metrics, such as throughput, for each UAV's data transmission across its flight path.}

\smallskip
\noindent {\bf {Industrial IoT} \/} \ \ \
{It is the network of interconnected smart devices, sensors, and machinery in industrial environments that collect, analyze, and act on data to optimize manufacturing, logistics, and other industrial operations~\cite{IIoT}.
WiFi-Diffusion enables the generation of fine-grained, high-quality radio maps, providing spectrum utilization analytics across spatial, temporal, and frequency domains within the industrial environment. 
These insights facilitate dynamic optimization of the placement and spectrum access strategies of smart devices, sensors, and machinery.
By implementing WiFi-Diffusion, critical industrial communication metrics, particularly latency and reliability, can be systematically enhanced, ensuring timely and robust data transmission for mission-critical industrial operations.}

\smallskip
\noindent {\bf {IoV} \/} \ \ \
{It is an intelligent ecosystem that connects vehicles, infrastructure, and users through Vehicle-to-Everything (V2X) communication, enabling real-time safety alerts, traffic optimization, and autonomous driving~\cite{IoV}.
The reliance on open wireless protocols makes the system susceptible to intentional jamming attacks and unintentional interference.
To mitigate these vulnerabilities, fine-grained, high-quality radio maps generated by WiFi-Diffusion can provide a critical defense mechanism by delivering dynamic spectrum awareness and enabling prediction of interference patterns across temporal, spatial, and frequency domains.
These radio maps endow IoV with anti-jamming capabilities and enable IoV to adaptively reconfigure communication strategies, e.g., proactive channel switching to uncontaminated frequencies. 
}

\section{Conclusion}\label{Conclusion}
Radio maps present communication parameters of interest, e.g., received signal strength, across a geographical region in a specific frequency band. 
It can be used to improve the efficiency of spectrum utilization, particularly critical for the WiFi spectrum.
The problem of fine-grained radio map estimation involves utilizing samples from sparsely distributed sensors to infer a high-resolution radio map.
This problem is uniquely challenging due to the ultra-low sampling rate, where the number of radio samples is significantly smaller than the high resolution required for radio map estimation.

In this paper, we design WiFi-Diffusion -- a novel generative framework for fine-grained WiFi radio map estimation using diffusion models.
WiFi-Diffusion consists of three blocks:
1) a boost block, which uses prior information such as the layout of obstacles and laws of radio propagation to optimize the diffusion model;
2) a generation block, which employs the diffusion model to generate a candidate set of fine-grained radio maps;
3) an election block, which utilizes the mathematical radio propagation model as a guide to find the best fine-grained radio map from the candidate set.
Extensive simulations show that 
1) WiFi-Diffusion constructs high-quality, fine-grained radio maps when the sampling rate is extremely low (less than $0.1\%$);
2) the radio map generated by WiFi-Diffusion can be an order of magnitude (10 times) better than that generated by SOTA when they share the same sampling rate;
3) WiFi-Diffusion requires only one-fifth of the sampling rate needed by SOTA to generate comparable radio maps.


\bibliographystyle{IEEEtran}
\bibliography{WiFiDiffusionRef}

\end{document}